\documentclass[conference]{IEEEtran}

\pdfoutput=1
\usepackage{tikz}
\usepackage{savesym}
\usepackage{amsmath, amssymb,amsthm, algorithm,algorithmic, tabularx}
\usepackage{cryptocode}
\usepackage{hyperref}
\let\labelindent\relax
\usepackage[inline,shortlabels]{enumitem}
\usepackage{pdfpages}
\usepackage{subfigure}
\usepackage{booktabs}
\usepackage{multirow}
\usepackage{amsmath}
\usepackage{sidecap}
\usepackage{makecell}
\usepackage{tabu}
\setlist{nolistsep}
\newtheorem{name}{Lemma}
\newtheorem{theorem}{Theorem}

\newtheorem{newdef}{Definition}
\usepackage{array}
\usepackage{graphicx}
\usepackage{etoolbox}
\usepackage{url}

\ifCLASSINFOpdf
\else
\fi
\begin{document}
%
\title{Crafter: Facial Feature Crafting against Inversion-based Identity Theft on Deep Models}

\author{\IEEEauthorblockN{Shiming Wang$^{1}$, Zhe Ji$^{1}$, Liyao Xiang$^{1\dag}$\thanks{$^{\dag}$Corresponding author.}, Hao Zhang$^1$, Xinbing Wang$^1$, Chenghu Zhou$^2$, Bo Li$^3$}
\IEEEauthorblockA{$^1$Shanghai Jiao Tong University, $^2$Chinese Academy of Science,  $^3$Hong Kong University of Science and Technology\\
$^1$\{my16wsm, ji\_zhe, xiangliyao08, mypeach, xwang8\}@sjtu.edu.cn, $^2$zhouchsjtu@gmail.com, $^3$bli@cse.ust.hk
}
}

\IEEEoverridecommandlockouts
\makeatletter\def\@IEEEpubidpullup{6.5\baselineskip}\makeatother
\IEEEpubid{\parbox{\columnwidth}{
    Network and Distributed System Security (NDSS) Symposium 2024\\
    26 February - 1 March 2024, San Diego, CA, USA\\
    ISBN 1-891562-93-2\\
    https://dx.doi.org/10.14722/ndss.2024.23326\\
    www.ndss-symposium.org
}
\hspace{\columnsep}\makebox[\columnwidth]{}}

\maketitle

\begin{abstract}
    With the increased capabilities at the edge (e.g., mobile device)  and more stringent privacy requirement, it becomes a recent trend for deep learning-enabled applications to pre-process sensitive raw data at the edge and transmit the features to the backend cloud for further processing. A typical application is to run machine learning (ML) services on facial images collected from different individuals. To prevent identity theft, conventional methods commonly rely on an adversarial game-based approach to shed the identity information from the feature. However, such methods can not defend against adaptive attacks, in which an attacker takes a countermove against a known defence strategy. 

    We propose Crafter, a feature crafting mechanism deployed at the edge, to protect the identity information from adaptive model inversion attacks while ensuring the ML tasks are properly carried out in the cloud.
    The key defence strategy is to mislead the attacker to a non-private prior from which the attacker gains little about the private identity.  
    In this case, the crafted features act like poison training samples for  attackers with adaptive model updates. Experimental results indicate that Crafter successfully defends both basic and possible adaptive attacks, which can not be achieved by state-of-the-art adversarial game-based methods.
    \end{abstract}
\section{Introduction}
\label{sec:intro}
Deep learning demonstrates impressive performance in many applications, owing to the complicated structures of learning models and massive crowdsourced data. Since local processing is often infeasible, the edge, e.g., mobile devices, collect the sensitive individual data, encode and transmit it to the untrusted cloud for further processing by learning models.
Facial image data raises the most concern as it is highly sensitive and susceptible to identity theft. Hence it is a critical issue to remove the sensitive identity information from the encoded features while accomplishing  cloud learning tasks. Examples could be makeup recommendations based on users' facial attributes or training facial expression detection model  on crowdsourced images, where identity information needs to be protected while preserving useful features.

Inversion attacks \cite{mahendran2014understanding,zhang2020secret,fredrikson2015model} can invert the private input pixel by pixel from the features, leading to identity leakage. The perception of identity involves not only small reconstruction distortion but also high-level semantic information. Thereby we propose \emph{identity perceptual privacy against inversion attacks}, which is  more complicated than reconstruction distortion-based defence \cite{xiao2020adversarial} and attribute inference-based  defence \cite{cherepanova2021lowkey,shan2020fawkes,li2020tiprdc,wang2021MI,li2021deepobf}. For instance, although \cite{cherepanova2021lowkey,shan2020fawkes} prevent  recognition models from inferring the identity attribute, their imperceptible perturbation fails to evade visual detection and still compromises privacy at the image level.

\begin{figure}[t]
	\centering
	\includegraphics[width=1\linewidth]{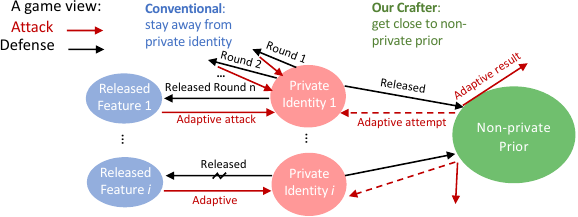}
	\caption{Conventional methods adopt a stay-away approach where the defence strategy is easily overwhelmed by an adaptive attacker step; our Crafter takes a get-close approach where the crafted features act like poison training samples to the adversary, disrupting the training of adaptive attackers. 
	}
	\label{fig:intuition}
\end{figure}

This motivates us to consider feature manipulation at the edge, so that transmitted features maintain high utility for the cloud ML task while protecting identity perceptual privacy. A naive solution is to train the learning model end to end to shed the identity information from the feature, which inevitably introduces a race between the defence party and the adversary \cite{xiao2020adversarial,li2020tiprdc,wang2021MI,li2021deepobf,singh2021disco}. In the race, the defence party pushes the feature away from the regime of private identity perception as in the left part of Fig \ref{fig:intuition}. The attacker could almost always overwhelm the defence party since the defence strategy is fixed and known upon the feature release, especially in the presence of adaptive attacks.

We aim to overcome this seemingly endless tit-for-tat between the defence party and the adversary by leveraging a non-private prior to bound the adversary's perceptual gain on private identities. Since the defender's strategy is fixed, its optimal move is to stay close to the adversary's prior (or a non-private prior in practice), thereby limiting the attacker's gain from the defender's move. 
As shown in the right part of Fig \ref{fig:intuition}, the adaptive attacker takes a countermove by mapping the non-private prior to different and independent identities, which disrupts the potential adaptive training of the attacker model. In addition, we assume the defender plays against a worst-case adversary, {\em i.e.,} a white-box attacker that obtains not only the defender's strategy but also all model weights to derive its move. If such an omniscient attacker can be defended, we have reasons to believe that the protection scheme is robust against other real-world attackers.

\begin{figure}[t]
	\centering
	\includegraphics[width=1\linewidth]{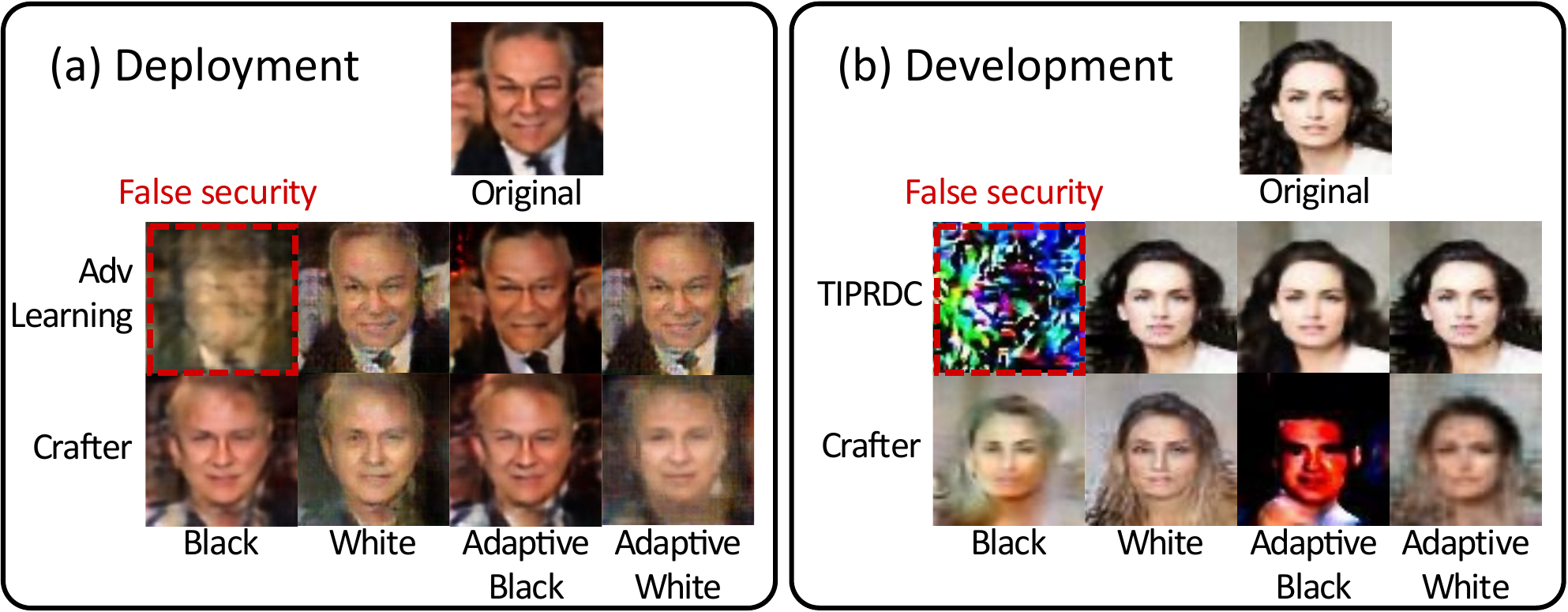}
	\caption{{Inversion results of existing defences and Crafter against multiple attacks. Crafter demonstrates robustness against both basic and possible adaptive attacks, while the baselines are not adaptive attacker-proof.}
	}
	\label{fig:visualization_one}
\end{figure} 

To this end, we propose Crafter, a facial feature crafting mechanism against inversion attacks on deep learning models. Given the original model, the framework perturbs the intermediate features to trick attackers into reconstructing non-private facial images while keeping the perturbation under a threshold to accomplish cloud tasks with high accuracies. 
What distinguishes Crafter from previous work is that it chooses not to \emph{erase} \cite{ragan2011decoupled,li2020tiprdc}, nor to \emph{obfuscate} \cite{dusmanu2020sift,chen2021perceptual} the concerned private attribute in feature representations, but to \emph{draw} the feature \emph{close to} a non-private prior  perceptually. The crafted features act as poison training samples to the inversion attacker, as they are close to the original features but drastically different in identity perception on the reconstructed image space. {Since neither image level distortion nor attribute level accuracy alone is sufficient to quantify Crafter's identity perceptual privacy, 
we propose a holistic privacy index, \emph{perceptual inversion indistinguishability}, as a distributional distance from the inversion attacker's prior to posterior view on the reconstructed images. We show through analysis and experiments that it is a valid privacy notion for Crafter's defence on facial images.}

Unlike the adversarial game-based methods, we demonstrate through analysis and experiments that Crafter consistently prevails over adaptive attackers that are specifically designed for Crafter. 
In addition, Crafter is able to decouple from the cloud learning tasks by taking advantage of the high-dimensional feature space and the robustness of deep models against minor input perturbations. It allows the feature to achieve the privacy goal with slight distortion, therefore not affecting the cloud inference or training performance. As the feature is expressed implicitly in our objective, we resort to the implicit function theorem to resolve the optimization challenge. Experimental results under various settings show  that Crafter successfully defends  black- and white-box attacks and their adaptive versions (Figure \ref{fig:visualization_one}), outperforming the state-of-the-art yet fulfilling the cloud tasks with high accuracies. 

In summary, our key contributions are as follows:
\begin{itemize}[leftmargin=*]
    \item 
	 We propose Crafter, a facial feature crafting approach that prevents identity leakage through inversion attacks, and is robust against possible adaptive attacks.
	\item
    
	We formulate the privacy of interest with perceptual inversion indistinguishability, a distributional distance between the attacker's posterior and prior beliefs on the reconstructed image space,  and show that Crafter achieves approximately optimal privacy-utility tradeoff.
	
    \item Crafter is open-sourced and easy to deploy as a plug-in to the edge-cloud computing framework, without any change in the backend models. \textbf{Code is available in the repository \cite{appendix}.}%
\end{itemize}

\section{Preliminaries}
The Earth Mover's distance (EMD) is a classic measure of inter-distribution distance, defined as
$
\operatorname{EMD}(P || Q)=\inf _{\gamma \in \Pi(P, Q)} \mathbb{E}_{(x, y) \sim \gamma}[\|x-y\|]
\label{eq:emd_original}$.  $P,Q$ denote distributions and ${\displaystyle \Pi (P,Q)}$ is the set of all joint distributions whose marginals are ${\displaystyle P}$ and ${\displaystyle Q}$. The infimum of the expectation is easy to compute for discrete tabular data,  but it is intractable to traverse all joints of  high-dimensional image distributions or continuous feature distributions. Hence we leverage its dual form:
 \begin{newdef}[KR duality of the Earth Mover's distance]
	 For distributions P and Q, and a 1-Lipschitz continuous function f, the EMD between the distributions is
	\begin{equation}
		\operatorname{EMD}(P||Q)=\sup _{\|f\|_{L} \leq 1} \mathbb{E}_{x \sim P}[f(x)]-\mathbb{E}_{x \sim Q}[f(x)].
	\end{equation}
\end{newdef}
\noindent In practice, we optimize a discriminator network $D$ to approximate the supremum on function $f$. This is a common practice in line with works on Wasserstein-GAN~\cite{arjovsky2017wasserstein,Gulrajani2017wgangp}. To encourage $D$ to be 1-Lipschitz,
i.e. $D$ has gradient with norm at most 1 everywhere, Gulrajani et al.
\cite{Gulrajani2017wgangp} enforce a gradient penalty term $g_p$ on the norm and adds it to the original EMD as a soft constraint: $g_p=\underset{\hat{x} \sim P_{\hat{x}}}{\mathbb{E}}\left[\left(\left\|\nabla_{\hat{x}} D(\hat{x})\right\|_{2}-1\right)^{2}\right]$, and $\hat{x}=\epsilon x_1 + (1-\epsilon)x_2 \sim P_{\hat{x}}$ is an interpolation of the two distributions where $x_1\sim P$ and  $x_2\sim Q$.

 We further show the key lemmas in Implicit Differentiation.
\begin{name}[{Cauchy, Implicit Function Theorem}]
	\label{lemma:IFT}
	For a function $f(x, y): \mathbb{R}^{n+m} \rightarrow \mathbb{R}^m$,
	if some 
	$(a,b)$ satisfies 
	\begin {enumerate*} [1) ]
	\item
	$
	f(a,b)=0$
	and 
	\item the Jacobian matrix $
	J_{f, {y}}({a}, {b})=\left[\frac{\partial f_{i}}{\partial y_{j}}({a}, {b})\right]
	$
	is invertible,
	\end {enumerate*}
	then surrounding $(a, b)$ there exist $U \subset \mathbb{R}^n$ and a unique continuously differentiable function $g:U \rightarrow \mathbb{R}^m$ that $g(a) = b$ and $f(x, g(x))=0, \forall x\in U$.	In addition,
	$\frac{\partial g}{x}(x)=-\left[J_{f, \mathbf{y}}(x, g(x))\right]^{-1}\left[\frac{\partial f}{\partial x}(x, g(x))  \right].$
\end{name}
\begin{name}[Lorraine~\cite{lorraine2020optimizing}, Neumann Inverse Approximation]
	\label{lemma:neumann}
	For a matrix $A$ and a sufficiently small scale $\alpha$ s.t. $|I- \alpha A |<1$
	, the following series holds and converges:
	$
	A^{-1}=
	\alpha \lim _{i \rightarrow \infty} \sum_{j=0}^{i}\left[I-
	\alpha A\right]^{j}.
	$
\end{name}
\section{Problem Setting and Threat Model}
\label{sec:threatmodel}
\subsection{Problem Setting}
\label{sec:probsetting}
We focus on the typical edge cloud computing scenario, where the cloud provides a crowdsourcing service that requires facial data from users to perform a joint machine learning (ML) task. Any raw data or feature transmission that potentially exposes  user identity is forbidden due to privacy concerns, and the cloud is responsible for providing
 a data collecting service to protect user privacy while accomplishing the crowdsourcing tasks.
We present two concrete examples for both \textit{model deployment} and \textit{model development} tasks to facilitate understanding. For model deployment, the service runs a trained  facial attribute-based makeup recommender to which users upload their face shots.
The recommeder analyzes the customers' facial characteristics but should not record their identities.
For model development, the service collects facial images from volunteers to train a facial expression detection model without distinguishing their identities.

\begin{figure}[t]
	\centering
 \includegraphics[width=0.8\linewidth]{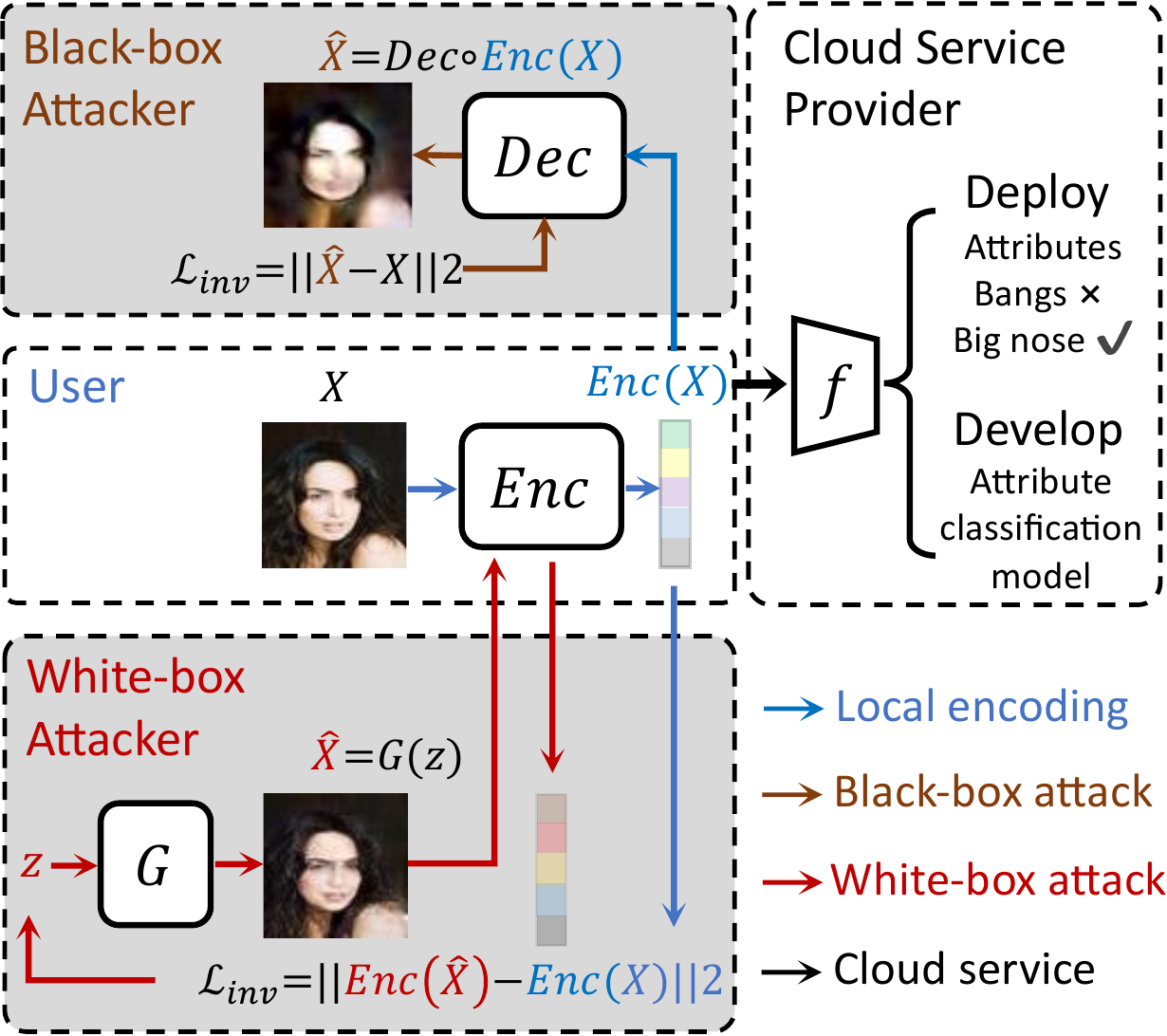}
	\caption{Problem overview. Users (blue) release locally encoded feature $Enc(X)$ of private image $X$ to complete computation tasks (black).
		Attackers intercept the released feature and attempt to reconstruct original private input through either black-box attack (brown) or white-box attack (red).
	}
	\label{fig:prob_set}
\end{figure}

To hide the identity information, the service deploys a local pre-processor to users and allow them to encode their images into features before transmission. 
An encoder $Enc$ extracts features from private images $X\in \mathcal{X}_{\mathrm{pvt}}$ and sends the features to the downstream computation task denoted by $f$. We formally show the problem setting in the lighted area of Figure~\ref{fig:prob_set}. The pipeline consists of an \emph{offline} and an \emph{online} stage involving three parties: \emph{user}, \emph{cloud service provider}, and \emph{adversary}. 
In the  deployment scenario, the service trains $Enc\circ f$ end to end offline, and distributes $Enc$ to the user while keeping $f$ at the cloud. Hence $Enc$ and $f$ are fixed  beforehand, and features $Enc(X)$ are fed into $f$ for online prediction tasks.  In the   development scenario, the service releases $Enc$ as a general feature extractor to the user  offline and collects $Enc(X)$, on which it runs the online training task for model $f$ which is not fixed apriori.   

\subsection{Threat Model}
\label{sec:threat_semantic}
Unfortunately, transmitting features still faces serious privacy risks.
     {We focus on non-targeted feature inversion attacks aiming to recover raw faces of unknown identities from the features.
   An inside-attacker in the untrusted cloud or a man-in-the-middle attacker receives the locally encoded raw feature $Enc(X)$   and  reconstructs    image $\hat{X}$.
 If $\hat{X}$ highly resembles the original face $X$  as shown in Fig \ref{fig:prob_set}   , the adversary would acquire the appearance of the  unknown user which clearly leaks user privacy: the adversary may link to an external database and decode the user's identity.   

  {Our threat model is different from those in certain existing defenses for user privacy. Instead of concrete attributes such as hair style and nose shape considered in 
\cite{ragan2011decoupled,li2020tiprdc,singh2021disco}, 
our adversary takes interest in user's identity, which is a semantic information particularly hard to isolate and remove from images.  In addition, we target inversion attacks that aim to unveil the initially undisclosed visage of a victim. We do not consider membership inference attacks as in 
\cite{wang2021datalens}, or attribute inference attacks as in
\cite{cherepanova2021lowkey,shan2020fawkes} 
where the adversary already obtains some of the victim's private facial images and attempts to infer the identity of its other faces with  recognition models. Detailed clarifications are in our online Appendix \ref{sec:appen_comparison} \cite{appendix}.
   To highlight this, we formally define the privacy of interest as    \emph{identity perceptual privacy against inversion attack}. We omit  ``against inversion attack" in the remainder of the paper for brevity}.

\noindent\textbf{{Identity perceptual privacy}}   describes the extent to which an inverted image is perceived  as the true private identity by an attacker, similar to a human observer’s interpretation. Obviously, this perceptual privacy cannot be simply characterized by the pixel-level distortion between $\hat{X}$ and $X$ (e.g., SSIM), or the  attribute inference accuracy (e.g., whether a recognition model extracts the correct ID from $\hat{X}$). For example, if $\hat{X}$ and $X$ are two images of the same identity taken under different light and angles, their SSIM is low but the recognition accuracy is still high; conversely, $\hat{X}$ can be visually similar to $X$ while evading facial recognition models, i.e. low accuracy but high SSIM. 
Typically, a high level of identity perceptual privacy suggests both  large image-wise distortion
 and 
low accuracy of 
facial recognition models.
To specify how an inversion attacker perceives an unknown identity, we instantiate reconstruction attacks with representative methods as in  Fig 3, categorized into basic and adaptive attacks according to whether an adversary adjusts its strategy to a given defence.

\noindent \textbf{Basic attacks} are generally categorized as either black-box or white-box.
A \textit{black-box adversary} queries the user's local encoder $Enc$ for unlimited times with  images of public identities    crawled from the Internet     {, denoted as $\mathcal{X}_{\text{pub}}$   .
It constructs a shadow decoder $Dec$ as the inverse of $Enc$ and trains the decoder
 as below \cite{black_inv}:
\begin{equation}
    \mathop{\min}_{Dec}~ \mathbb{E}_{X' \in \mathcal{X}_{\mathrm{pub}}} \|Dec(Enc(X')) - X'\|_2.
    \label{eq:black}
\end{equation} 
We denote the reconstructed image as $\hat{X}^*=Dec(Enc(X)).$

In contrast, a \textit{white-box adversary} has unrestricted access to $Enc$ and its parameters. 
The adversary first trains a Wasserstein-GAN to distill public prior knowledge of general facial images from public datasets $\mathcal{X}_{\mathrm{pub}}$ \cite{zhang2020secret}.
Given latent vectors   $z_r$    following random distributions, the pretrained generator $G$ is able to generate realistic-looking facial images  with no particular private identity, referred to as \emph{average faces}.
Initiated from this public prior, 
the adversary reconstructs a target image via gradient-based optimization on latent representation $z${, starting from some random $z_0$}:
\begin{equation}
    z^*=\arg\mathop{\min}_{z} \|Enc(X) - Enc\circ G(z)\|_2,
    \label{eq:za-loss}
\end{equation}
and the reconstructed image is 
$
    \hat{X}^* = G(z^*).
$
We refer to it as a canonical white-box attacker, and $z^*$ as the \emph{best-response} latent representation of image $X$'s feature.
To avoid confusion, we use $z_{\mathrm{org}}$ to denote the best-response of the raw feature $Enc(X)$.
 {An et. al~\cite{key} further propose a more advanced white-box reconstruction with potentially higher fidelity using StyleGAN instead of the canonical WGAN.}

Alternatively, a white-box adversary can initialize its optimization with a black-box decoder, referred to as a \textit{hybrid white-box adversary}. Upon receiving  $Enc(X)$, the attacker initializes the image as output of the pre-trained black-box $Dec$ and runs pixel-level optimization to minimize feature loss: 
\begin{equation}
    \mathop{\min}_{\hat{X}} \|Enc(X)-Enc(\hat{X})\|_2,~ \hat{X}_{\mathrm{init}}=Dec(Enc(X)).
    \end{equation}
Notice that white-box access is a realistic assumption in our scenario. As discussed in \S\ref{sec:probsetting}, the cloud distributes $Enc$ to users. Thus an inside attacker naturally has white-box access to $Enc$, and a man-in-the-middle adversary can disguise itself as a benign user and acquire the parameters.

\noindent \textbf{Adaptive attacks.} Apart from the basic attacks discussed above, adversaries can specifically adjust their strategies to target a given defense. As an example, the adversary may update $Dec$ and $G$ with protected features $F_X$.
  Intuitively, such adaptive attacks can be stronger as they leverage the   protection strategy    and try to bypass it. We consider all possible adaptive approaches under our edge cloud scenario. 
Detailed definitions of the adaptive attacks are deferred to \S\ref{sec:false}, after we introduce our protection mechanism.

\noindent \textbf{Attacker's knowledge.}
We assume the attackers, basic or adaptive, have access to any public datasets crawled from the Internet, and any $Dec$ and $G$ models. There is no constraint on the adversary's reconstruction approach. The attacker is free to choose between   decoder-based    ($Dec$), GAN-based ($G$) and more advanced StyleGAN-based methods.
   {The adversary is assumed to have no access to the private images of the unknown identity, i.e., $\mathcal{X}_{\text{pub}}$ and $\mathcal{X}_{\text{pvt}}$ has no identity overlapping. Intuitively, if the adversary already acquires multiple faces of a victim, the harm is already done and it is meaningless to prevent another image from exposure. Hence the adversary cannot launch an attribute inference attack on the features directly: it can only train the  identity classifier on $\mathcal{X}_{\text{pub}}$ which does not effectively recognize the identities of $\mathcal{X}_{\text{pvt}}$. 
We will further extend the attacker's capability in \S\ref{sec:extended}.}

\noindent \textbf{{Defender's knowledge.}}   {We assume the defender has knowledge of the white-box attacker's reconstruction loss function $\mathcal{L}_{\text{inv}}$, which is the $L_2$ feature distortion between the original and the reconstructed as in most inversion attacks. 
It is later verified that our defence using $\mathcal{L}_{\text{inv}}$ is also effective against black-box and other attacks. The defender also has full access to a trained $G$ provided by trusted third parties, which can be any open platform offering widely-acknowledged pretrained generators. To be noted, the defender knows nothing about the adversary's reconstruction model, indicating the adversary does not necessarily use $G$ in its attack.}}

\section{Methodology}
\label{sec:methodology}
In face of the privacy threats, the cloud provides a user encoder to let users craft features with the following goals:
\begin{itemize}[leftmargin=*]
\item \textit{Privacy}: given features, the attacker reconstructs images that reveal little private identity information.
\item\textit{Utility}: the crafted features should complete the subsequent ML tasks with high performance.
\item \textit{Robust against adaptive attacks}: once features are released, attackers cannot bypass the defense via adaptive updates.
\end{itemize}

Towards these goals, we illustrate our design choices in \S\ref{sec:motivation} and our approach of privacy-preserving feature representation construction in \S\ref{sec:overview}. The key idea is to iteratively craft a feature $F_X$ to bring the adversary's posterior close to a non-private prior (minimizing $\mathcal{L}_p$) while restricting feature perturbation (minimizing $\mathcal{L}_u$). To capture the privacy leakage, we give a theoretical privacy guarantee for our method in \S\ref{sec:priv notion}. Finally, we discuss the advantage of Crafter in real-world cases in \S\ref{sec:opt}.
 Notations used are listed in Table~\ref{tab:notation}.

\begin{table}[t]
  \centering
  \small
    \caption{Notations.}
      \resizebox{1\columnwidth}{!}{
    \begin{tabularx}{0.5\textwidth}{l|X}
      \small
      \textbf{Notation} & \textbf{Definition}\\
      \hline
      \multicolumn{2}{c}{DNN models:} \\
      \hline
      \emph{Enc} & The local encoder under attack.\\
      \emph{G}& Pretrained generator; input: latent vectors; output: images.\\
      \emph{D}& Discriminator that attempts to distinguish reconstructed images from  attacker's prior. \\
      \hline
      \multicolumn{2}{c}{Variables:} \\
      \hline 
      $X$ & Private input images.\\
      $F_X$& Protected feature representation of $X$.\\
      $z_r$ & Latent random vector.\\
      $z_{\mathrm{org}}$ & Best-response latent vector of raw feature $Enc(X)$.\\
      $z^*(F_X)$& Best-response latent vector of feature $F_X$.\\
      
      $G(z_r)$ 
      &  Prior belief.\\
      $G(z^*(F_X))$ &Posterior belief given $F_X$.\\
      \hline
      \multicolumn{2}{c}{Loss functions:} \\
      \hline 
      $\mathcal{L}_p$& Privacy loss,  EM distance between the distributions of reconstructed and prior images.  \\
      $\mathcal{L}_u$& Utility loss,  deviation from the protected feature to the original feature. \\
      $\mathcal{L}_{\mathrm{inv}}$& Reconstruction loss, deviation from features of reconstructed images  to  original features.
    \end{tabularx}%
      }
     \label{tab:notation}
\end{table}%

\subsection{Design Choices}
\label{sec:motivation} 
We make the following design choices according to the goals above:

\noindent \textbf{Feature-manipulation protection.}
We manipulate the locally encoded feature representation $Enc(x)$ before release so that the attacker fails to extract private information from what it intercepts. The reason that we do not perform image-level manipulation~\cite{cherepanova2021lowkey,shan2020fawkes} is that previous image perturbation either fails to prevent white-box attack, or is visually identifiable (See \S\ref{sec:eval}). Further, we do not replace the local encoder $Enc$ with other models since $Enc$ has been pre-trained to fit the downstream tasks. A simple replacement may fail to meet the inference or training requirement. 

\noindent\textbf{Protection against white-box attacks.}
In our setting, $Enc(\cdot)$ is deployed by the service beforehand and thus can be acquired by the adversary. Our scheme should fight against a white-box attacker which is typically stronger than a black-box attacker. A defence method that withstands the stronger white-box attacker suffices to transfer well to the weaker black-box inversion attacks. 

\noindent\textbf{Exploiting non-private prior.}
To achieve the privacy goal, previous work formulates a multi-player game: the defender maximizes the difference between information revealed by released features and the private raw images; the attacker updates itself simultaneously against defender's strategy~\cite{li2020tiprdc,xiao2020adversarial}. We refer to the type of strategy as the \emph{stay-away} (from the original) approach. 
However, there is no guarantee that the attacker strategy is worst-case at the end of the optimization. Hence the attacker can proceed the adversarial game given a fixed defence strategy and eventually undermines the protection.

In contrast to the stay-away approach, we choose a \emph{get-close} method to draw the released features close to a public prior on the reconstructed image space. Notice that we make no assumption on the defender's knowledge about any particular adversary's prior: any public prior will suffice as long as it does not overlap users' private information. 
Ideally, the exposed features do not enhance the attacker's knowledge.
They act like poison training samples to the inversion attacker, as they are close to the raw data on the feature space but drastically different on the inverted image space. 
Hence the attacker will only corrupt its model if it adaptively updates its strategy based on these poison samples, thus breaking the adversaial game.

\noindent\textbf{Distributional distance as privacy loss.}
To better quantify what an attacker perceives from the reconstruction, we follow the well-defined `perceptual quality' formulation in signal restoration 
\cite{rdp,perceptual,blau2018perception} 
and adopt distributional distance as our privacy-preserving loss. 
In image signal restoration, perceptual quality refers to how much an output signal $\hat{X}$ is perceived by humans as a realistic sample. It quantifies the perception of `naturalness' with the distance between the distribution of output signals and the distribution of natural signals, eg. EMD 
\cite{blau2018perception,rdp}. Similarly, we quantify the attacker's perception of identity using the EMD between the possible reconstruction distribution and the non-private average face distribution. Just as a smaller EMD in signal restoration indicates the output signal is perceptually closer to realistic images, a smaller EMD in our scenario implies that the reconstruction is more likely to be perceived as non-private images and thus higher identity perceptual privacy.

\begin{figure}[t]
	\centering
	
 \includegraphics[width=0.75\linewidth]{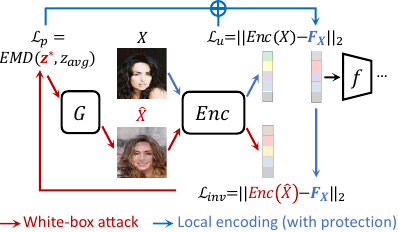}
	
	\caption{Overview of our feature crafting scheme against inversion attack. Attacker (red) obtains a best-response latent vector $z^*$ of protected feature $F_X$ by minimizing inversion loss $\mathcal{L}_{\mathrm{inv}}$. Defender (blue) manipulates $F_X$ to balance privacy $\mathcal{L}_{\mathrm{p}}$ of reconstructed $z^*$ and utility $\mathcal{L}_{\mathrm{u}}$ for computation tasks.}
	\label{fig:our_prt}
\end{figure}

\subsection{Privacy-Preserving Feature Crafting}
\label{sec:overview}
Our privacy-preserving feature construction scheme embeds carefully-crafted perturbations in the feature representation before releasing it to complete the computation tasks. Such a perturbation is crafted to disrupt the attacker's reconstruction ability via feature collision~\cite{ali2018poison}, misleading the attacker's view of the private input to some non-private prior, and is kept small to maximally retain the utility of the downstream tasks.  

We show the design overview in Figure~\ref{fig:our_prt}. Given private images $X$, the defender   simulates a white-box attacker's reconstruction result $\hat{X}^*=G(z^*)$ by minimizing the inversion loss $\mathcal{L}_{\mathrm{inv}}$. $\hat{X}^*$ is used to measure the adversary's perception of certain identity. To prevent the identity leakage from $F_X$, the defender brings the distribution of reconstructed images close to that of the average faces, which are not associated with any private identity   ({\em i.e.}, minimizing $\mathcal{L}_{p}$). Meanwhile, the perturbation on feature should be limited so that the computation tasks are not disrupted ({\em i.e.}, minimizing $\mathcal{L}_u$). We will demonstrate each part in the following.

\noindent\textbf{Privacy protection.}
    Following the design choice, we describe attacker's identity perception by the EMD between the distributions of attacker's inverted images and prior belief $G(z_r)$. The \emph{privacy loss} is:
\begin{equation}
    \label{eq:loss p}
    \mathcal{L}_p(z^*(F_X))= \text{EMD}(G(z^*(F_X))||G(z_r)),
 \end{equation}
  where $z^*(F_X)$ is the best-response latent representation of $F_X$ that minimizes the white-box inversion loss
$
\mathcal{L}_{\mathrm{inv}}(F_X, z)=\left\|F_{X}-Enc \circ G\left(z\right)\right\|_2$ which replaces $Enc(x)$ in Eq.~(\ref{eq:za-loss}) with $F_X$. As discussed in \S\ref{sec:motivation}, $\mathcal{L}_p$ represents the enhancement of attacker's knowledge given $F_X$ compared to its prior belief of general facial images.
With this loss minimized, the attacker is tricked into generating close-to-average faces and perceiving   non-private identities, thereby achieving a successful defence.

\noindent\textbf{Utility preservation.}
We restrict $F_X$'s deviation from the original feature $Enc(X)$ to {prevent severe drops in downstream utility.} 
The \emph{utility loss} is:
\begin{equation}
    \mathcal{L}_u(F_X)= \left\|F_X - Enc(X)\right\|_2.
\end{equation}
Notice that the utility loss does not concern the downstream model $f$. Given an encoder that already functions well for $f$ (either pretrained in the deployment scenario or generally provided in the development scenario), $f$ is expected to be robust under minor deviation from its original input $Enc(X)$, {and intuitively the larger the feature deviation, the larger the impact on utility}. This independence of $f$ decouples privacy goals from the downstream model, and thus admits unknown computation tasks.

\noindent\textbf{Overall objective.}
Given a target network $Enc \circ f$, a private input $X\in \mathcal{X}_{\mathrm{pvt}}$ and a generator $G$ trained with public images $\mathcal{X}_{\mathrm{pub}}$, 
the high complexity and nonlinearity of \emph{Enc} and \emph{G} makes it possible to find a feature $F^*_X$ that collides with the original feature $Enc(X)$, 
while its best-response reconstructed image $G(z^*)$ approximates to the average face distribution $G(z_r)$ in image space. The overall goal is to seek an ideal spot in the utility-privacy tradeoff. Hence the optimization objective is such that 
\begin{gather}
    \label{eq:overall}
    \min_{F_X} \mathcal{L}_p(z^*)~ \text{where} ~ z^{*} =\arg \min _{z}\mathcal{L}_{\mathrm{inv}}(F_X, z), \\
    \text{subject to}~ \mathcal{L}_u(F_X) \leq  l . \notag
\end{gather}

We adopt its Lagrange dual form, and transform the minimization of EMD (Eq. \ref{eq:loss p})) which has no closed-form solution to a minimax game as the canonical Wasserstein-GAN. This formulation utilizes neural network, so we use `neural net distance' $d_{nn}(\cdot, \cdot)$ with EMD interchangeably. Thereby we formulate our protection scheme as:
\begin{gather}
    \label{eq:ultimate}
    \min_{F_X}  \max_{|D|_L \leq 1}\mathcal{L}_p(D, z^*(F_X)) +  \beta \cdot \mathcal{L}_u(F_X) ~\text{where}\\
    z^{*}(F_X) =\arg \min _{z}\mathcal{L}_{\mathrm{inv}}(F_X, z), \notag  \\
 \begin{aligned}
    &\mathcal{L}_p(D,z^*(F_X))= \mathbb{E}_{z_r}[D\circ G(z_r)] -\mathbb{E}_{z^*}[D\circ G(z^*(F_X))] \notag ,\\
    &\mathcal{L}_u(F_X)= \left\|F_X - Enc(X)\right\|_2 \notag ,\\
    &\mathcal{L}_{\mathrm{inv}}(F_X, z)=\left\|F_{X}-E n c \circ G\left(z\right)\right\|_2. \notag
\end{aligned} 
\end{gather}
The discriminator $D$ introduced here attempts to distinguish the reconstructed image from the average face. It is different from the pretrained $D$ in the white-box attack (Eq.~(\ref{eq:pretrain}) in the appendix \cite{appendix}). {Also notice that the pretrained $G$ is fixed, and it is the feature $F_X$ that competes with the discriminator $D$.}

The above formulation is a nested optimization. Solving it with gradient-based optimizers is challenging as one must differentiate through the best-response latent vector $z^*$ as a function of the feature $F_X$.  To address this problem, we propose a method based on the Implicit Function Theorem (IFT) to compute the privacy loss gradient with respect to $F_X$. 

\noindent\textbf{Optimization via IFT.}
\label{sec:prt_ift}
We show how to solve Eq.~(\ref{eq:ultimate}) to seek an optimized $F^*_X$. 
$\frac{\partial \mathcal{L}_p(D, z^*)}{\partial D}$ and $\frac{\partial \mathcal{L}_u(F_X)}{\partial F_X}$ are both \emph{direct gradients}, and can be directly computed. The bottleneck lies in the \emph{indirect gradient} $\frac{\partial \mathcal{L}_p(D, z^*)}{\partial F_X}$, since $z^*$ changes in each iteration with respect to the protected feature. $\frac{\partial z^*(F_X)}{F_X}$ is difficult to obtain as $z^*$ is determined by optimizing $\mathcal{L}_{\mathrm{inv}}(F_X, z)$. We thus resort to the IFT (Lemma~\ref{lemma:IFT})  
and compute the indirect gradient $\frac{\partial \mathcal{L}_{p}(D,z^*(F_X))}{\partial {{F_X}}}$ as
\begin{equation}
    \label{eq:dp_df}
    -\alpha \frac{\mathcal{L}_{p}(D,{z^*})}{\partial z^*} 
    \cdot
    \lim _{i \rightarrow \infty} \sum_{j=0}^{i}\left[I-
    \alpha\frac{\partial^{2} \mathcal{L}_{\mathrm{inv}}}{\partial z \partial z}\right]^{j}
    \cdot
    \frac{\partial^{2} \mathcal{L}_{\mathrm{inv}}}{\partial z \partial {F_X}} \notag
\end{equation}
   Detailed derivations are in Appendix \ref{appen:ift} \cite{appendix}. Having tackled the implicit differentiation, we are ready to solve Eq. \eqref{eq:ultimate}. 

\subsection{Algorithm of Crafter}
\label{appen:algs}
We outline our scheme in Alg.~\ref{alg:opt}. Alg.~\ref{alg2} is adopted from \cite{lorraine2020optimizing} for computing the indirect gradient $\frac{\partial \mathcal{L}_{p}(D,z^*)}{\partial {{F_X}}}$. 
In Alg.~\ref{alg:opt}, the simulated attacker intercepts the feature representations of a batch of private images $X \in \mathbb{R}^{b\times (w\times h)}$, and computes the corresponding $z^* \in \mathbb{R}^{b\times d}$. Each $z^{*(j)}, j\in \{1,\cdots,b\}$
generates a reconstructed image $G(z^{*(j)}) \in \mathbb{R}^{w \times h}$, which can be considered as a sample from the distribution of reconstructed images, rather than the distribution of pixels in the image. The discriminator respectively samples $m$ images from the reconstructed images and the average images, trying to distinguish the two groups of data at each iteration of optimization. Unlike the canonical WGAN where a generator directly competes with the discriminator, our generator $G$ is pretrained on public images and fixed during the process. It is the feature $F_X$ that strives to confuse the discriminator. We follow the tradition in WGAN that the discriminator undergoes multiple training steps (lines 5 to 8) for each update of $F_X$.
\begin{algorithm}[htbp]
	\renewcommand{\algorithmicrequire}{\textbf{Input:}}
	\renewcommand{\algorithmicensure}{\textbf{Output:}}
	\caption{Crafter}
	\label{alg1}
	\begin{algorithmic}[1]
		\label{alg:opt}
		\REQUIRE Target network \emph{Enc}, generator \emph{G}, a batch of private images \emph{X} of batch size \emph{b}, minibatch size \emph{m}, latent vector dimension \emph{d}, training iterations of the discriminator per feature update $n_{\mathrm{critic}}$, tradeoff scale $\beta$.
		\STATE Initialization: $F_X \leftarrow Enc(X)$,    $z_{\mathrm{avg}} \leftarrow \texttt{randn}(b, d)$
		  $z_r \leftarrow \texttt{randn}(b, d)$
		 \WHILE{$F_X$ has not converged}
		\STATE  $z^{*} =\arg \min _{z}\mathcal{L}_{\mathrm{inv}}(F_X, z)$
		\FOR{$t=0,\dots, n_{\mathrm{critic}}$}
		\STATE Sample $\{z^{*(j)}\}_{j=1}^m$ a minibatch from inverted $z^*.$
		\STATE Sample $\{z_r^{(j)}\}_{j=1}^m$   a minibatch from random $z_r$ .
		\STATE $\mathcal{L}_p \leftarrow 
		\frac{1}{m} \sum_{j=1}^{m} \left[
		D_{\omega}\circ G(z^{*(j)})
		-D_{\omega}\circ G(z^{(j)}_{\mathrm{avg}})\right]+g_p$
		\STATE $\omega \leftarrow \mathrm{\texttt{AdamOptimizer}}(\nabla_{\omega}\mathcal{L}_p, \omega)$
		\ENDFOR
		\STATE
		  $\mathcal{L}_p \leftarrow \frac{1}{bs}\sum_{j=1}^{b}-D_{\omega}\circ G(z^{*(j)})$
		 \STATE $v_1 \leftarrow \texttt{approxInverseHVP}(\frac{\partial \mathcal{L}_p}{\partial z^*}, 
		\frac{\partial \mathcal{L}_{\mathrm{inv}}}{\partial z^*}
		)$
		\STATE $v_2 \leftarrow \beta\frac{\partial{\mathcal{L}_u}}{\partial F_X} - \texttt{grad}(
		\frac{\partial \mathcal{L}_{\mathrm{inv}}}{\partial z^*}, F_X, \texttt{grad\_outputs}=v_1
		)$
		\STATE $F_X \leftarrow \mathrm{\texttt{AdamOptimizer}}(v_2, F_X, lr=flr)$
		\ENDWHILE
		 \STATE  $F^*_X \leftarrow F_X$  
		 \ENSURE  Crafted feature $F^*_X$.
	\end{algorithmic}  
\end{algorithm}

\begin{algorithm}[htbp]
	\renewcommand{\algorithmicrequire}{\textbf{Input:}}
	\renewcommand{\algorithmicensure}{\textbf{Output:}}
	\caption{approxInverseHVP$(\frac{\partial \mathcal{L}_p}{\partial z}, 
		\frac{\partial \mathcal{L}_{\mathrm{inv}}}{\partial z})$. \\Experiments used the default values $\alpha=0.001$, $si=150$}
	\begin{algorithmic}[1]
		\label{alg2}
		\STATE Initialization: $p \leftarrow \frac{\partial \mathcal{L}_{p}}{\partial z}$
		\FOR{$j=0,\dots, i$}
		\STATE $v \leftarrow v-\alpha \cdot \texttt{grad}(\frac{\partial \mathcal{L}_\mathrm{inv}}{\partial z}, z, \texttt{grad\_outputs}=v)$
		\STATE $p \leftarrow p + v$
		\ENDFOR
		\ENSURE $\alpha p$
	\end{algorithmic}  
\end{algorithm}

To sum up,  our feature crafting system  operates in the following  two phases.

\noindent \textbf{Offline:} $G$ and $Enc$ preparation. 
 The trusted party collects a public image dataset $\mathcal{X}_{\mathrm{pub}}$ on which it trains a WGAN following Eq.(\ref{eq:pretrain}), and releases the trained generator $G$. $G$ takes in random latent vectors and outputs realistic-looking facial images. The user receives $G$ from   any  trusted party and the local encoder model $Enc$ from cloud depending on the 
 utility tasks. 
  For a deployment task, $f$ is also readily deployed on cloud. For a development task, features from users are crowdsourced for training new models on cloud.

\noindent\textbf{Online:} feature crafting and task completion.
 After determining $Enc$ and $G$, the user runs Alg.\ref{alg:opt} to construct $F_X$ of its private images and sends $F_X$ to the cloud. 
The server receives $F_X$. If it undertakes a deployment  task, the features go through model $f $ to return a prediction result. Otherwise, the server collects features as training data to train a target model.

\subsection{  $\epsilon$-Perceptual Inversion Indistinguishability  }
\label{sec:priv notion}

 We formally define $\epsilon$-perceptual inversion indistinguishability (PII) which is inspired by the concepts of differential privacy and $t$-closeness. The PII is directly defined on the EMD loss which indicates how a simulated white-box attacker perceives identities from reconstructions.  Specifically, let $\mathcal{X}_{\text{pub}}$ be the public dataset with no private identity involved. For any feature $F_X$, let $G\circ F_X$ denote the inverted distribution using white-box $G$, i.e. $G \circ F_X = G(z^*(F_X))$. We then have:

\begin{newdef}
	[$\epsilon$-Perceptual Inversion Indistinguishability]
	\label{def:RI}
	A feature crafting system $M$ is $\epsilon$-perceptual inversion indistinguishable ($\epsilon$-PII) on the private image $\mathcal{X}_{\text{pvt}}$ if
	\begin{equation}
		\label{eq:PII}
		\operatorname{EMD}(G\circ M(\mathcal{X}_{\text{pvt}})|| G\circ M(\mathcal{X}_{\text{pub}})) \leq \epsilon
	\end{equation}
 \end{newdef}

 The interpretation of Def.~\ref{def:RI} is that the smaller the privacy upperbound $\epsilon$, the closer $M$ is to ideal privacy perceptually against the inversion attack, and thus the less identity leakage.
Note that PII is not associated with any realistic attacker but only a public generator $G$ serving as a simulated privacy indicator
 --- in fact the specific choice of $G$ would  affect only the value of $\epsilon$, but not the qualitative relationship between $\epsilon$ and the realistic defence capability. That is, as long as $G$ is well-trained (eg. provided by a trusted third party and capable of reliable inversions), the defence strength against realistic attacks will increase as $\epsilon$ decreases.

  Crafter meets the definition with $G\circ M(\mathcal{X}_{\text{pvt}})$ being $G(z^*(F^*_X))$ for $X \in \mathcal{X}_{\text{pvt}}$, and $G\circ M(\mathcal{X}_{\text{pub}})$ being the non-private prior $G(z_r)$. Hence as Crafter optimizes the feature $F_X$ towards minimizing $\text{EMD}(G(z^*(F_X))||G(z_r))$ in Eq. (\ref{eq:loss p}), it minimizes the left-hand-side of Eq. (\ref{eq:PII}), thereby satisfying Def.~\ref{def:RI} with a smaller $\epsilon$ indicating stronger perceptual privacy.

\noindent \textbf{Validity of $\boldsymbol{\epsilon}$-PII.}
We show $\epsilon$-PII is a valid indistinguishability index by providing 1) discussion on avoiding potential limintation of PII; 
2) experimental consistency of perceptual privacy in \S\ref{sec:inference}. 
A potential limitation of the $\epsilon$-PII formulation is that EMD may not measure the inter-distribution distance precisely when the stability difference between the users' reconstructed $M(\mathcal{X}_{\text{pvt}})$ and the public prior $ M(\mathcal{X}_{\text{pub}})$ is too large. That is, the $\epsilon$-value might not reflect the true dissimilarity between the distributions.
To reduce the stability difference, Crafter normalizes the public faces according to the pixel mean and variance of users' private images so that 
the reconstructed images are likely to be pixel-level-similar to the public faces $G(z_r)$ in Def.~\ref{def:RI}.

\noindent\textbf{Connection to other privacy concepts.}
We notice that $\epsilon$-PII is related to the conventional $\epsilon$-differential privacy and $t$-closeness definitions. We analyze their relations as below.
\begin{newdef}
	[{Adjacent datasets}] Two datasets $D$ and $D'$ are adjacent if they differ in the existence of a single user's data.
\end{newdef}
\begin{newdef}[{$\epsilon$-Differential Privacy}]
	Let $D_{\infty}(P||Q)$ denotes the max divergence between distributions $P$ and $Q$.
	A randomized mechanism $M$ is $\epsilon$-differentially private ($\epsilon$-DP) if its distribution over any two adjacent datasets $D$ and $D'$ satisfies $D_{\infty}(M(D)||M(D')) \leq \epsilon$.
\end{newdef}
DP and PII share the same intuition: bounding the impact that the private data's presence has on the mechanism outputs.
The key differences lie in their definitions of paired datasets, the dependency of $\epsilon$ values, and the choices of divergence function.
DP considers a pair of adjacent dataset $(D,D')$ differing on a single user's private data record, while the $\mathcal{X}_{\text{pvt}}$ and $\mathcal{X}_{\text{pub}}$ considered in PII differ on the existence of the private identities, which aligns with our goal of quantifying identity leakage.
Second, $\epsilon$ in DP is a privacy upper bound on any adjacent sets, focusing on the worst-case privacy of the mechanism itself on any possible data record.
 In contrast, PII is contingent on the given private data $\mathcal{X}_{\text{pvt}}$, emphasizing the privacy of the user data under the mechanism's protection. PII is also dependent on the public generator $G$ quantitatively but not qualitatively as shown by our experiments.
Finally, PII adopts EMD rather than the max divergence $D_{\infty}$ for the inter-distribution divergence, as the EMD is closely related to human perception rather than the theoretical worst case. 

Next we compare our proposed $\epsilon$-PII to $t$-closeness, which is conventionally applied to structural tabular data.
\begin{newdef}[{Equivalence class}]
	Let quasi-identifiers be attributes whose values when taken together can potentially identify an individual in an anonymized data table.
	An equivalence class of a data table is a set of records that have the same values for the quasi-identifiers.
\end{newdef}
\begin{newdef}[{The $t$-closeness principle}]
	An equivalence class has t-closeness if the EMD between the distribution of a sensitive attribute $S$ in this class and the distribution of the attribute in the whole table is no more than a threshold $t$.
\end{newdef}
PII and $t$-closeness both limit the knowledge gain between the prior and posterior view of the attacker (or observer in the tabular data context).
We list their counterparts for comparison in Table \ref{tab:tcloseness}. The major differences are as follows.
The entity that $t$-closeness aims to protect is an equivalence class in the data table, while for $\epsilon$-PII is the private raw images as a whole.
The private information at risk for $t$-closeness is a sensitive attribute (eg. a disease), while for PII it is the reconstructed images. Finally, the public prior information assumed by $t$-closeness is the whole datatable distribution,
whereas in  PII it is the public images with no identity overlapping.

	\begin{table}[ht]
		\centering
		\caption{Comparison of $t$-closeness and $\epsilon$-PII.}
		\label{tab:tcloseness}
		\begin{tabular}{c|c|c}
			\toprule
			Privacy notion &{$t$-closeness} & $\epsilon$-PII \\
			\midrule
			Protected entity &{Equivalence class} & Private raw images \\
			Attacker's goal&{Sensitive attribute $S$} & Reconstructed images \\
			\cmidrule{1-3}
			\multirow{2}{*}{Prior}&Distribution of S in  & Distribution of reconstructed \\
			&the whole datatable&public images \\
			\cmidrule{1-3}
			\multirow{2}{*}{Posterior}&Distribution of S in & Distribution of reconstructed  \\
			&the equivalence class &private images \\
			\bottomrule
		\end{tabular}
	\end{table}
	Finally, we provide a theoretical upper bound on $\epsilon$ of Crafter in Appendix~\ref{appen:proof} \cite{appendix}. The theoretical result indicates that at the same utility loss, Crafter offers an $\epsilon$ within a bounded distance to the infimum $\epsilon$, thereby achieving an approximate optimal privacy-utility tradeoff.

\section{Discussion}
\label{sec:opt}
We answer the following questions in this section: 
\emph{
	Does Crafter remain robust against an adaptive attacker? Why does Crafter use implicit optimization?}

 \subsection{Robustness against Adaptive Attacks}
\label{sec:false}

 We explore three possible adaptive attacks  against Crafter.
Details of the design idea are in Appendix \ref{appen:adaptive_idea} \cite{appendix}.
Experimental results are in \S\ref{sec:adaptive}.

\noindent\textbf{A1: Continue the optimization.}
Once the protection is completed, the feature $F_X^*$ is released and is fixed ever since. 
We design A1 that continues to optimize its attack model against the protection.
Specifically for Crafter, A1 queries Crafter with its own images $X$, intercepts the corresponding protected features $F_{X}^*$, and obtains the reconstructed image $\hat{X}=G(z^*(F_X^*))$ (white-box A1) or $\hat{X}=Dec(F_X^*)$ (black-box A1).
Then it updates $G$  or $Dec$  to minimize the reconstruction loss $\mathcal{L}_{\text{attacker}}=\|\hat{X}-X\|_2$. 
The same can be done for other adversarial game-based defences  \cite{xiao2020adversarial,li2020tiprdc}.
We establish the family of adversarial game-based defences, and show why they are vulnerable to A1 while Crafter remains secure.

\begin{newdef}[{The family of adversarial game-based protection.}]
	\label{def:future-proof}
	Given a defencer $\mathcal{P}$ with strategy $x_1$ and an attacker $\mathcal{A}$ with strategy $x_2$, a game-based protection framework is:
	\begin{gather}
		\label{eq:game}
		\mathrm{Find}~x^*=(x_1^*, {x_2}^*) ~\mathrm{s.t.} \\
		\begin{align}
			x_1^* &=\arg \min _{x_1}\mathcal{L}_{\mathrm{privacy}}(x_1,x_2^*)+\beta \cdot \mathcal{L}_{\mathrm{utility}}(x_1)\notag\\
			x_2^* &=\arg\min_{x_2}\mathcal{L}_{\mathrm{attacker}}(x_1^*, x_2, X_{\mathrm{test}}). \notag
		\end{align}
	\end{gather}
\end{newdef}
\begin{table*}[t]
	\centering
	\small
	\caption{List of game-based defence. $CE$ denotes the cross-entropy loss. $C$ classifies features to different private attributes $u$. $I( ; )$ means mutual information.}
	  \small
	\resizebox{\linewidth}{!}{
	\begin{tabular}{
			p{0.11\linewidth}
			p{0.09\linewidth}p{0.06\linewidth}
			p{0.22\linewidth}
			p{0.20\linewidth}
			p{0.20\linewidth}}
		\hline
		Name & $x_1$ of $\mathcal{P}$ & $x_2$ of $\mathcal{A}$ & $\mathcal{L}_{\mathrm{privacy}}$ & $ \mathcal{L}_{\mathrm{utility}}$ & $\mathcal{L}_{\mathrm{attacker}}, X\in X_{\mathrm{train}}$\\
		\hline
		Adv Learn~\cite{xiao2020adversarial} & \emph{Enc,~f} & \emph{Dec}
		&$-\|\emph{\text{Dec}}\circ\emph{\text{Enc}}(X) - X\|_2$
		&$CE(f\circ \emph{\text{Enc}}(X), Y)$
		&$\|\emph{\text{Dec}} \circ \emph{\text{Enc(X)}} - X\|_2$\\
		{Disco}~\cite{singh2021disco} & {\emph{Enc,~f,~Pruner}} & {\emph{Dec}} & {$-||\emph{\text{Dec}}\circ \emph{\text{Pruner}}\circ \emph{\text{Enc}}(X)-X||$} &	{$CE(f\circ \emph{\text{Pruner}}\circ \emph{\text{Enc}}(X),Y)$} & {$||\emph{\text{Dec}}\circ \emph{\text{Pruner}}\circ \emph{\text{Enc}}(X)-X||$}
		\\
		TIPRDC~\cite{li2020tiprdc} & \emph{Enc} & \emph{C}
		&$-CE(\emph{\text{C}}\circ \emph{\text{Enc}}(X), u)$
		& $I(\emph{\text{Enc}}(X); X)$
		&$CE(\emph{\text{C}}\circ \emph{\text{Enc}}(X), u)$\\
		\hline
		Ours & $F_X$ & {\emph{G}}
		&$\mathrm{EMD}(G(z^*(F_X)), G(z_r))$
		& $\|F_X-\emph{\text{Enc}}(X)\|_2$
		& {$\|G(z^*(F_X))-X\|_2$ }\\
		\hline
	\end{tabular}}
	 \label{tab:family}
\end{table*}

Table~\ref{tab:family} describes our and previous defence \cite{xiao2020adversarial,li2020tiprdc,singh2021disco}  under the above framework. 
{$\mathcal{A}$ aims to minimize its loss $\mathcal{L}_{\text{attacker}}$ on the private test set $X_{\text{test}}$, which is not available to $\mathcal{A}$, so in practice $\mathcal{A}$ will use $X_{\text{train}}$ instead.}
Previous work commonly adopt an update approach which converges at $(\hat{x}_1,\hat{x}_2)$, and claim the protection successful as $\hat{x}_1$ has an advantage over $\hat{x}_2$.
However, none of them guarantees the adversary $\hat{x}_2$ at convergence is the worst-case.
An adaptive $\mathcal{A}_1$ can thus continue to optimize $x_2$ and reduce the attacker loss on $X_{\mathrm{train}}$, which is the opposite of the privacy loss under the \emph{stay-away} approach. Therefore, the adaptive attacker successfully undermines the defence and transfers well to $X_{\mathrm{test}}$. 

Our framework is free of this worry. We argue that our simulated $\mathcal{A}$ without any adversarial update is stronger than any potential A1 adaptive attacks $\mathcal{A}_1$. Under the \emph{get-close} approach, $\mathcal{A}$ is misled to reconstruct 
images close to random $G(z_r)$, as shown in Figure \ref{fig:tsne}.
$\mathcal{A}_1$ essentially matches $G(z_r)$ with $X$, which are independent and identically distributed image samples.
Establishing correspondence between independent samples only results in larger $\mathcal{L}_{\mathrm{attacker}}$ than $\mathcal{A}$ and weakens $\mathcal{A}_1$.
The same holds for black-box adaptive attacks. Hence the simulated attacker $\mathcal{A}$ is stronger than any $\mathcal{A}_1$, and the attacker is discouraged from adversarial updates. If $\mathcal{P}$ has an advantage over $\mathcal{A}$, 
it is  robust against A1 adaptive attacks.
 \begin{figure}
	\centering
	\includegraphics[width=0.7\columnwidth]{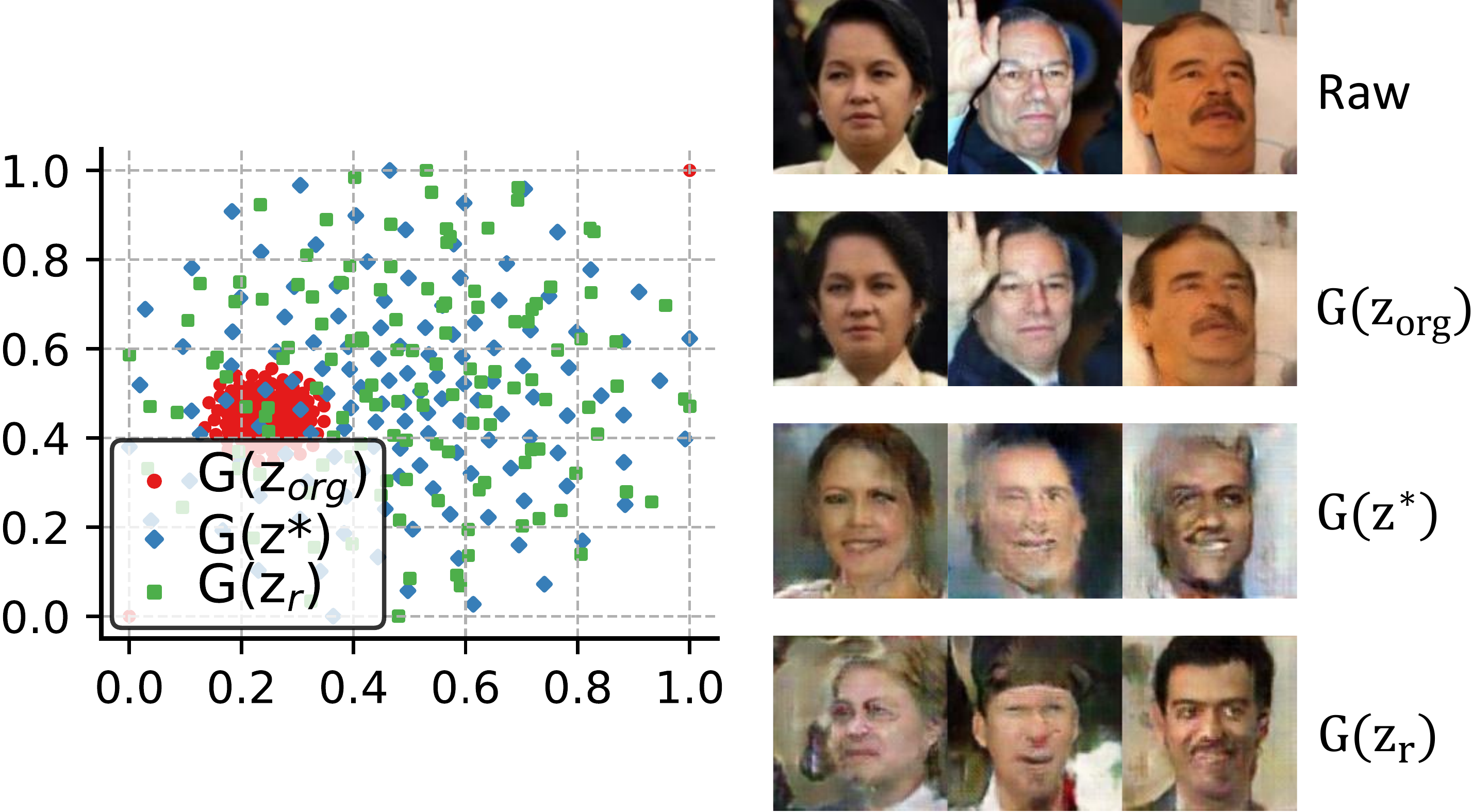}
	 \caption{Visualization of the best-response  $z_{\mathrm{org}}$ of the raw feature (red), $z^{*}$ of our crafted feature (blue), and  {$z_r$} the attacker's prior (green). Our framework shifts the unprotected posterior belief towards attacker's prior belief.}
	\label{fig:tsne}
\end{figure}

\noindent\textbf{A2: Utilize different generators.}
Crafter's optimization relies on a specific simulated generator model $G$.
This adaptive adversary uses generator models that are different from and possibly stronger than $G$.
Specifically, we evaluate our scheme on generators of different structures and latent dimensions, including the more advanced StyleGAN as proposed in~\cite{key}, and show through experiments (\S\ref{sec:adaptive}) that Crafter is robust against different generators. This is because Crafter transfers well to different A2 adaptive attacks, as long as the simulated attack used in training can reliably extract private identity information through reconstruction. Hence, when choosing the simulated $G$ in Eq.~\eqref{eq:ultimate}, a generator model with fair reconstruction performance on $Enc$ is sufficient for qualified defence across different A2 adaptive attacks. 
A2 attacks may also leverage $G$ trained with different public datasets.
However, we empirically discover that the attack is usually the strongest when the adversary uses the same public dataset for its WGAN, so we show the worst-case results and only present results using different public sets for StyleGAN as in \cite{key}.

{\noindent\textbf{A3: Average features over multiple queries.}} 
This adaptive adversary waits for a user to query Crafter on the same batch of image multiple times, and averages over the multiple protected features before reconstruction.
Privacy of Crafter is solely accomplished by feature  perturbation. 
As discussed in \S\ref{sec:threatmodel}, a user can run defence on the same batch of image multiple times and ends up with different feature perturbations because of the randomness of $z_r$ in Eq. (\ref{eq:ultimate}) during each iteration.
 As a result, the perturbation of each query may offset each other, and averaging over the queries may remove the perturbation.

We show in \S\ref{sec:adaptive} that this averaging strategy indeed undermines the defence. However, by shuffling user's batch of data each time feeding into Crafter, we ensure robustness against A3 adaptive attacks.
We implement shuffling as an inherent part of the encoder in Crafter, so that no additional abnormal query detection is required for the user.
 The intuition is that A3 works only when image batches of different queries are identical, including the order of images. If the attack averages the features of two batches containing the same images but in shuffled orders, features of different images are mismatched. Perturbations can be hardly removed if the input batches contain a sufficient number of images. If a batch merely has a few images, Crafter records the features the first time the batch is fed, and reuses the features afterwards to prevent averaging.

\noindent\textbf{False security} means a protection lacks  important robustness   evaluations against comprehensive adaptive attacks \cite{tramer2020adaptive}.
 A protection must be effective against realistic adaptive attacks to be of practical use.
 The adaptive attacks in this section could be launched without additional assumptions, and
 are all easy to implement in real applications.
 Therefore, if a protection fails to defend adaptive attacks, privacy against basic attacks is meaningless and is merely a false security. No matter how strong the privacy is, the adversary can always
 breach the security with a simple adaptive approach.

\subsection{Implicit Optimization}
\label{sec:prt_z}
One may question the necessity of using implicit optimization in \S\ref{sec:prt_ift}. Indeed, we can evade the indirect gradient computation if we perform optimization on the latent representations instead of manipulating features. Specifically, in \S\ref{sec:prt_ift}, we control the latent $z$ instead, and the protected feature is a function of $z$: $F_X = Enc\circ G(z)$. The optimization on latent $z$ thus becomes:
\begin{gather}
    \label{eq:zp}
    \min_{z}  \max_{|D|_L \leq 1}\mathcal{L}_p(D, z) +  \beta \cdot \mathcal{L}_u(z) ~\text{where}\\
    \begin{aligned}
    &\mathcal{L}_p(D,z)= \mathbb{E}_{z_r}[D\circ G(z_r)] -\mathbb{E}_{z}[D\circ G(z)] \notag ,\\
    &\mathcal{L}_u(z)= \left\|Enc\circ G(z) - Enc(X)\right\|_2. \notag 
\end{aligned}
\end{gather}
Since $F_X$ can be computed by forwarding $z$ through NNs, we can apply gradient-based optimizers on $z$ and $D$. The algorithm can be found in Appendix~\ref{sec:crafter-z} \cite{appendix}. 
We refer to this alternative as \emph{Crafter-z}. 
Although it evades implicit differentiation, it delivers poor privacy-utility tradeoff empirically, mostly because the tradeoff is better to manipulate in the feature space ($F_X$) rather than the latent space ($z$). We will elaborate on this in \S\ref{sec:inference}.

\section{Evaluation}
\label{sec:eval}
We aim to answer the following questions in this section:
\begin{enumerate}[start=1,label={\bfseries Q\arabic*:}]
    \item Is Crafter effective against white-box attacks? 
    \\Does Crafter transfer well against black-box attacks?
    \\How well does Crafter maintain downstream utility?
	\item{{Is Crafter robust against the three adaptive attacks?}}
    \item What is the advantage of using implicit optimization? 
    \item Does Crafter introduce large runtime overhead?
\end{enumerate}

\begin{figure*}[ht]
	\centering
	\includegraphics[width=7in]{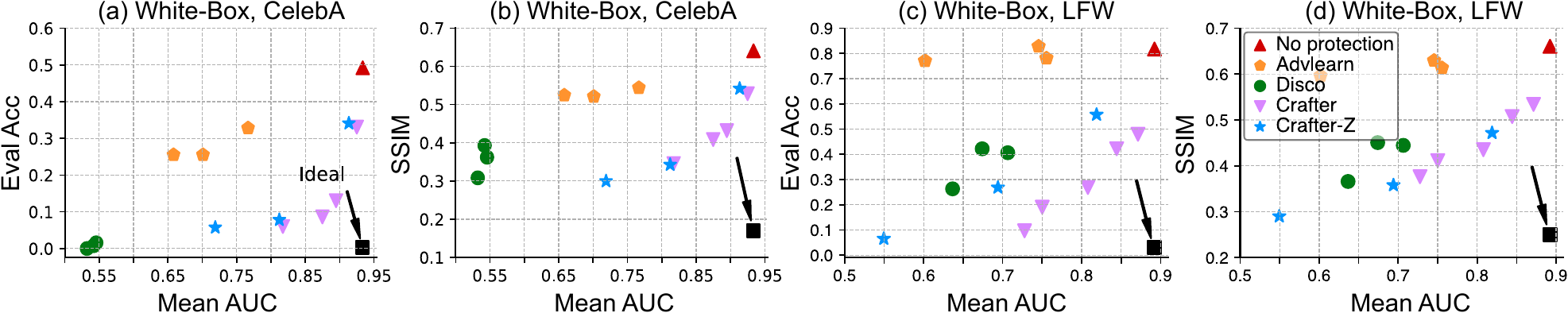}
	\vspace{-3mm}
	\caption{Privacy-utility tradeoffs against white-box attacks. On CelebA, $\beta \in \{0.5,1,2,10\}$ for Crafter, and $\beta \in \{20, 50\}$ for Crafter-z. On LFW, $\beta\in \{3.5, 4, 4.5, 6, 7\}$ for Crafter, and $\beta\in \{5, 10, 20\}$ for Crafter-z. Subfigures share the same legend. {For Adv learning, $\lambda \in \{0.1, 0.5, 0.8\}$. For Disco, $\lambda \in \{0.2, 0.6, 0.8\}$.} The black square denotes the ideal tradeoff point.}
	\label{fig:scatter_white}
\end{figure*}

\subsection{Setup}
\label{sec:setup}
\noindent\textbf{Implementation.}
We implement Crafter with {\tt PyTorch 1.10.0} and run all experiments on NVIDIA GeForce RTX 3090 GPU. We first act as the trusted party to train $G$ on $\mathcal{X}_{\mathrm{pub}}$. The cloud trains $Enc$ and $f$ end to end on $\mathcal{X}_{\mathrm{pub}}$ in the {deployment} scenario, or leverages a general feature extractor $Enc$ in the {development} scenario. Users collect $Enc$ and $G$ to generate crafted features of $\mathcal{X}_\mathrm{test}$, which accomplish the subsequent downstream tasks.

\noindent\textbf{Datasets.} 
We use the widely-adopted CelebA~\cite{wang2016celeba}, LFW~\cite{liu2015LFW} and VGGFace2 for training and testing Crafter. CelebA is labeled with 40 binary facial attributes, and is split into 200K images for public set $\mathcal{X}_{\mathrm{pub}}$, 17K images for private train set $\mathcal{X}_{\mathrm{train}}$ and 4K for private test set $\mathcal{X}_{\mathrm{test}}$. The input dimension of each image is 64$\times$64. For LFW, we choose 10 independent binary facial attributes and split the dataset into 10K images for $\mathcal{X}_{\mathrm{pub}}$, 2K for private train $\mathcal{X}_{\mathrm{train}}$ and the rest 1K for private test $\mathcal{X}_{\mathrm{test}}$. We crop and resize each image to 128$\times$128. The public $\mathcal{X}_{\mathrm{pub}}$ has no identity overlapping with the private sets, while $\mathcal{X}_{\mathrm{train}}$ and $\mathcal{X}_{\mathrm{test}}$ is a 4:1 (2:1) split for each private identity's images in CelebA (LFW). For VGGFace2, we crop and rezie images to 112$\times$112, and perform 2:1 train-test split. Note that on each dataset, $\mathcal{X}_{\mathrm{train}}$ is used in baselines, or to train the oracle evaluating networks, not by Crafter. We consider `identity' as the private attribute to be protected whereas the cloud tasks are 40 facial attributes classification, 10 attributes classification, and a 5-class hair color classification, for CelebA, LFW, VGGFace2, respectively.

\noindent\textbf{Models.}
For the target models under attack, we use the classic image processing DNNs (ResNet18, VGG16 and ResNet50) as $Enc \circ f$ in the {deployment and development} scenarios. $Enc$ is chosen as the first few layers of the models. For $D$ in Crafter and in the white-box attacker model, a CNN model is adopted. We prepare three generator models --- $G_1, G_2$ and StyleGAN (\cite{key}) --- for white-box attacks, and a decoder $Dec$ for black-box attacks.  $G_1, G_2, Dec$ are composed of stacks of \texttt{ConvTranspose2D} layers. For StyleGAN, $\mathcal{X}_{\mathrm{pub}}$ is from CelebA, and $\mathcal{X}_{\mathrm{test}}$ is from VGGFace2 following the design in \cite{key}; for other models, $\mathcal{X}_{\mathrm{pub}}$ and $\mathcal{X}_{\mathrm{test}}$ are from the same dataset.
For the evaluating networks, we adopt ResNet152 for CelebA, Facenet~\cite{schroff2015facenet} for LFW, and {Azure Face API~\cite{azure} for VGGFace2}. Detailed architecture of the networks is in Appendix~\ref{sec:architecture} \cite{appendix}. 

\noindent\textbf{Metrics.} We use the mean AUC as the utility metric to evaluate the performance of cloud tasks. For privacy, we simulate white-box, black-box, hybrid white-box and adaptive attacks to reconstruct images from intercepted features. Hyperparameters of the attacks are in Appendix~\ref{sec:attacker_param} \cite{appendix}.
{The empirical identity perceptual privacy of each defence is evaluated against the attacks by the following metrics:} 
\begin{itemize}[leftmargin=*]
\item\textit{Evaluation Accuracy (Eval Acc).}
 We use a face verification model as well as the Microsoft Azure Face {API~\cite{azure}} as the evaluating networks, trained on $\mathcal{X}_{\mathrm{train}}$. The evaluation accuracy is the identification accuracy on the reconstructed private test images. 
\item\textit{Feature Similarity (FSIM).}
We  feed the inverted and raw private $\mathcal{X}_{\mathrm{test}}$ into the evaluating network, extract the penultimate layer outputs and calculate their cosine similarity. 
\item\textit{SSIM}  evaluates the resemblance between the reconstructed images and the original ones on a pixel level. It is a supplement of the semantic security metrics above.
\item\textit{Human study.} We conduct a human study to further quantify Crafter's privacy performance following the design in~\cite{key}. 
\end{itemize}
Different from simply applying the evaluation accuracy of a specified private attribute, our combination of privacy metrics goes beyond attribute-level or pixel-level privacy, but evaluates identity privacy, as mentioned in \S\ref{sec:threat_semantic}.

\noindent\textbf{Baselines.}
We compare Crafter with state-of-the-art privacy-preserving approaches against inversion attacks.
\textit{Adv Learning}~\cite{xiao2020adversarial} and \textit{Disco} \cite{singh2021disco} fall under the model deployment scenario.
\textit{TIPRDC}~\cite{li2020tiprdc} falls under the development scenario.
\textit{Fawkes}~\cite{cherepanova2021lowkey} and \textit{LowKey}~\cite{shan2020fawkes} are image-manipulation-based defences  against attribute inference attack, and we adapt them to our threat model.
We implement Crafter-z as a supplementary baseline as discussed in \S\ref{sec:prt_z}. 

\begin{itemize}[leftmargin=*]
    \item \textit{Adv Learning}~\cite{xiao2020adversarial} presents an adversarial game-based approach by pitting the encoder \emph{Enc} and downstream \emph{f} against a black-box decoder $Dec$ trained on $\mathcal{X}_{\mathrm{pub}}$. 
	The tradeoff hyperparameter $\lambda \in \{0.1, 0.5, 0.8\}$. 
	{It is a task-oriented protection, which requires prespecified utility tasks and downstream network $f$. 
	Therefore, this is a baseline under the model development scenario.}
	{\item \textit{Disco} \cite{singh2021disco} takes the same adversarial game-based approach as Adv Learning~\cite{xiao2020adversarial}, but it further introduces a pruner to mask privacy-leaking feature channels. We set its tradeoff parameter as $\lambda \in \{0.2, 0.6, 0.8\}$. It is also a baseline under the model deployment scenario.
	}
    \item \textit{TIPRDC} aims to protect private attribute (ID in our setting) through  adversarially training \emph{Enc} on $\mathcal{X}_{\mathrm{train}}$ to minimize the mutual information between the feature and the input. The tradeoff hyperparameter are chosen as $\lambda \in \{0.1, 0.5, 0.8\}$.
    It is a task-independent protection, which generates feature representataion of raw input images for unknown downstream tasks. Hence it falls into the category of baselines under the model development scenario.
    \item \textit{Fawkes} \& \textit{LowKey} methods~\cite{cherepanova2021lowkey,shan2020fawkes} perturb the input images under Fawkes mode \{low, mid, high\} (or iterations $\in$ \{50, 75, 100\} for LowKey) to mislead an ensemble of ID classifiers. Their original design is to release the perturbed training images as poison samples for the adversarial ID classifiers. To fairly compare with Crafter, we replay their approach on the private test set to see how it preserves the input privacy.
\end{itemize}

\subsection{Defence against Basic Attacks}
\label{sec:inference}

\noindent\textbf{{Model deployment scene.}}
We show the Crafter's performance in comparison with baselines in the deployment scenario. For fair and integral comparison, we report results under a variety of utility-privacy tradeoffs by tuning the hyperparameter $\beta$. A discussion on $\beta$ can be found in the end of this section. The ideal tradeoff is a point with a high AUC and a low privacy loss metric value, as depicted by the black square in each figure.

\textit{White-box attacks.}
As Figure~\ref{fig:scatter_white} shows, Crafter gives the best tradeoff against white-box attacks in almost all cases, as it is closest to the lower-right corner (\textbf{Q1}). We leave FSIM-utility plots to {Appendix~\ref{sec:FSIM} \cite{appendix}} as FSIM is mostly consistent with Eval Acc. In Figure~\ref{fig:scatter_white}(a), $\beta=1$ reduces the Eval Acc of the unprotected from 49.22\% to as low as 8.59\%, while downstream AUC drops a mere 0.05. In contrast, Crafter-z offers a less satisfactory tradeoff under the white-box attack. For example, on a 5.7\% reconstructed Eval Acc, the average AUC of Crafter-z drops to 0.72, supporting the use of implicit optimization other than direct optimization (\textbf{Q3}). Among the three baselines, Adv Learning fails to defend against white-box attacks on both datasets, {\em e.g.,} on CelebA, the Eval Acc is still high around 30\% while it sacrifices 0.20 AUC.
{Disco improves the tradeoff upon Adv Learning on LFW, but is still inferior to Crafter as in Figure \ref{fig:scatter_white}(c)(d).
On CelebA, Disco achieves strong privacy but low utility akin to random guessing (around 0.55 AUC) for all tradeoff parameter $\lambda$ values (Figure \ref{fig:scatter_white}(a)(b)).
}
 We test Crafter and the baselines on Microsoft Azure Face and obtain similar tradeoffs. The results are in Appendix~\ref{sec:append_inference} \cite{appendix}.

\textit{Black-box attacks.}
We test Crafter and baselines against the black-box decoder $Dec$ trained on the feature-image pairs of the public data. Comparison between Fig~\ref{fig:scatter_white} and Fig~\ref{fig:scatter_black} illustrates that white-box attackers are generally stronger than black-box ones,  {\em e.g.,} the 50\% Eval Acc in Figure~\ref{fig:scatter_white}(a) is higher than the 34\% in Figure~\ref{fig:scatter_black}(a). And Crafter achieves a comparable or lower value of Eval Acc or SSIM against black-box attacks, and obtains satisfactory utility-privacy tradeoff (\textbf{Q1}). On CelebA, Figure~\ref{fig:scatter_black} show that Crafter-z achieves a comparable tradeoff to Crafter, but is not a proper alternative to the IFT-based Crafter, as the tradeoffs are mostly manipulated through the learning rate $lr_{z}$ other than $\beta$ in Crafter-z (\textbf{Q3}, further discussion in Appendix~\ref{sec:crfz} \cite{appendix}). 

\begin{figure}[ht]
	\centering
	\includegraphics[width=3.5in]{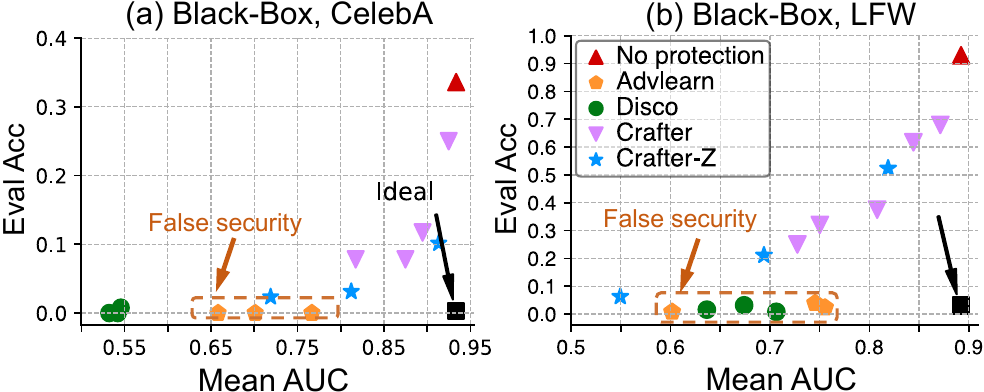}
	\vspace{-3mm}
	\caption{Privacy-utility tradeoffs against black-box attacks, {deployment scenario}. For Crafter, $\beta \in \{0.5,1,2,10\}$ on CelebA and $\beta\in \{3.5, 4, 4.5, 6, 7\}$ on LFW. For Crafter-z, as $\beta$ cannot properly trade off privacy and utility, we manipulate the learning rate instead: $lr_{z} \in \{0.0001,0.0005,0.001\}$ on CelebA and $lr_{z} \in \{0.0001, 0.001, 0.01\}$ on LFW.}
	\label{fig:scatter_black}
\end{figure}

\begin{figure}[ht]
    \centering
    \includegraphics[width=3.5in]{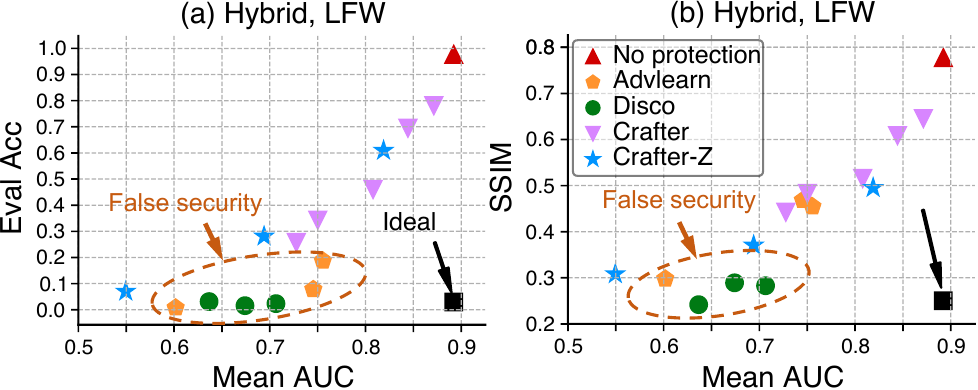}
    \caption{Privacy-utility tradeoffs against hybrid attacks, {deployment scenario}.}
    \label{fig:scatter_hybrid}
    \end{figure}


\textit{Hybrid white-box attacks.}
Since white-box inversion starts off from some random $z$ which can affect optimization, it does not triumph black-box attackers pervasively. This is exactly the case for LFW: without any protection, the EvalAcc of white-box attacks is 81.77\%, while that of black-box is high as 92.97\%. Hence we initiates a hybrid white-box attack which starts from the output of the black-box attack. The hybrid one successfully outperforms with an EvalAcc of 97.65\%. Figure~\ref{fig:scatter_hybrid} shows that Crafter gives better tradeoffs when pitting against this hybrid attacker than other baselines except for Adv Learning and Disco. 

\textit{False security.}
Adv Learning and Disco display much higher robustness against black-box and hybrid attacks than Crafter, shown by Figure~\ref{fig:scatter_black} and \ref{fig:scatter_hybrid}. However, we point out that it is an unreliable \textbf{false security} discussed in \S\ref{sec:false}, as an A1 adaptive attack could easily break them. \S\ref{sec:adaptive} will show that Crafter is in fact superior to the baselines against black-box and hybrid attacks.

\begin{figure*}[ht]
    \centering
    \includegraphics[width=7in]{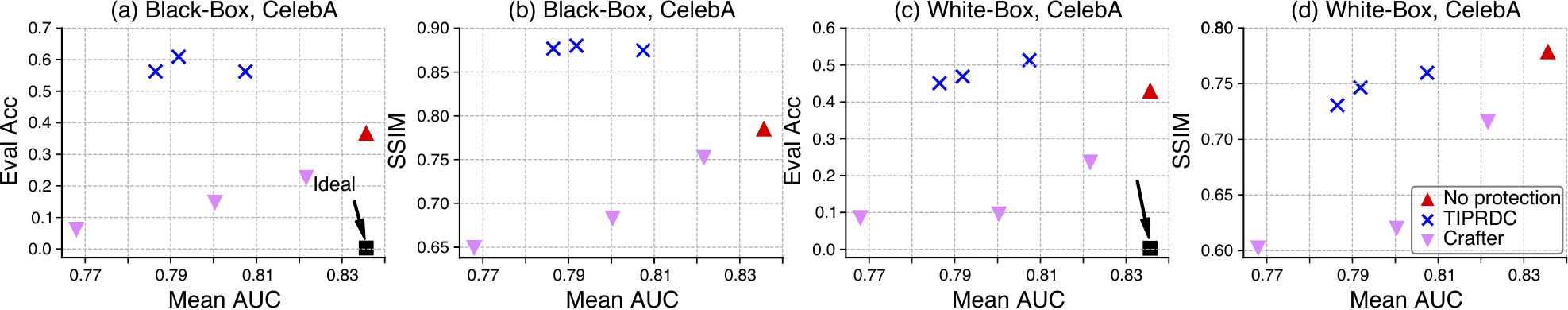}
    \caption{Crafter and TIPRDC on CelebA against white/black-box attacks, deployment scenario. $\beta \in \{0.5,1,5\}$ for Crafter, and $\lambda \in \{0.1, 0.5, 0.8\}$ for TIPRDC.}
    \label{fig:scatter_training}
    \end{figure*}

\noindent\textbf{Model development scene.}
We compare the performance of Crafter with TIPRDC on CelebA against white-box and black-box attacks, and report the privacy-utility tradeoffs across $\beta$s. The cloud task is to train a 40-facial-attribute classifier given features of $\mathcal{X}_{\mathrm{train}}$, with its performance evaluated  by the mean AUC on $\mathcal{X}_{\mathrm{pub}}$. As in Figure~\ref{fig:scatter_training}, Crafter gives a better tradeoff against both white- and black-box attacks than TIPRDC. At $\beta=1.0$,  Crafter reduces the EvalAcc of the unprotected from 43\% to 10\%, while AUC drops a mere 0.02. Moreover, privacy loss metrics of black-box attacks is close to white-box attacks, confirming the transferability of Crafter against attacks. 

In contrast, TIPRDC fails to preserve privacy against attacks as all choices of $\lambda$ lead to an Eval Acc above 40\%. For a fair analysis, we directly report evaluation results against adaptive attacks. As we analyze, TIPRDC exhibits good privacy performance in \cite{li2020tiprdc} on the binary sensitive attribute (\emph{e.g.,} gender), which is easy to isolate from the insensitive semantic information. 
Our setting requires a higher level of semantic privacy, to preserve the identity information that depend on the general appearance and is hard to separate from the input. Hence erasing such private information is contradictory to TIPRDC's utility goal of maximally preserving the semantic information of raw inputs. The conflict goal of utility and privacy disrupts the TIPRDC encoder, yielding features containing ample original information yet fails in the downstream tasks. This explains why the Eval Acc of TIPRDC is even higher than that of the unprotected feature.

A naive baseline in this scenario is to perform the training task using only $\mathcal{X}_{\text{pub}}$, which provides perfect privacy as no private images are involved. However, it may exhibit unsatisfactory utility due to potentially imbalanced utility labels for the unknown training tasks. We simulate an imbalanced $\mathcal{X}_\text{pub}$ by randomly choosing 1 among 40 binary utility attributes in CelebA and removing the images with the attribute labeled '0'. The utility AUC of model trained on $\mathcal{X}_\text{pub}$ alone drops to 0.76, while that of $\mathcal{X}_\text{pub} \cup \mathcal{X}_\text{pvt}$ can reach 0.82. Hence users' $\mathcal{X}_\text{pvt}$ is needed to augment $\mathcal{X}_\text{pub}$ to improve model performance.

\noindent\textbf{Impacts of hyperparamter $\boldsymbol{\beta}$.} 
We take a closer look at how $\beta$ gauges the tradeoff between privacy and utility.
Regradless of the attacker type, data points of Crafter in Fig \ref{fig:scatter_white}, \ref{fig:scatter_black}, \ref{fig:scatter_hybrid} from left to right correspond to $\beta=0.5, 1, 2, 10$ on CelebA and $\beta=3.5, 4, 4.5, 6, 7$ on LFW under the deployment scenario.
A decreasing $\beta$ allows a less informative image to be reconstructed but undermines the cloud utility. The same is true for the development scenario in Figure \ref{fig:scatter_training}.
Hence Crafter's tradeoff is easy to manipulate by a single coefficient $\beta$. 
\begin{figure*}[ht]
    \centering
    \includegraphics[width=7in]{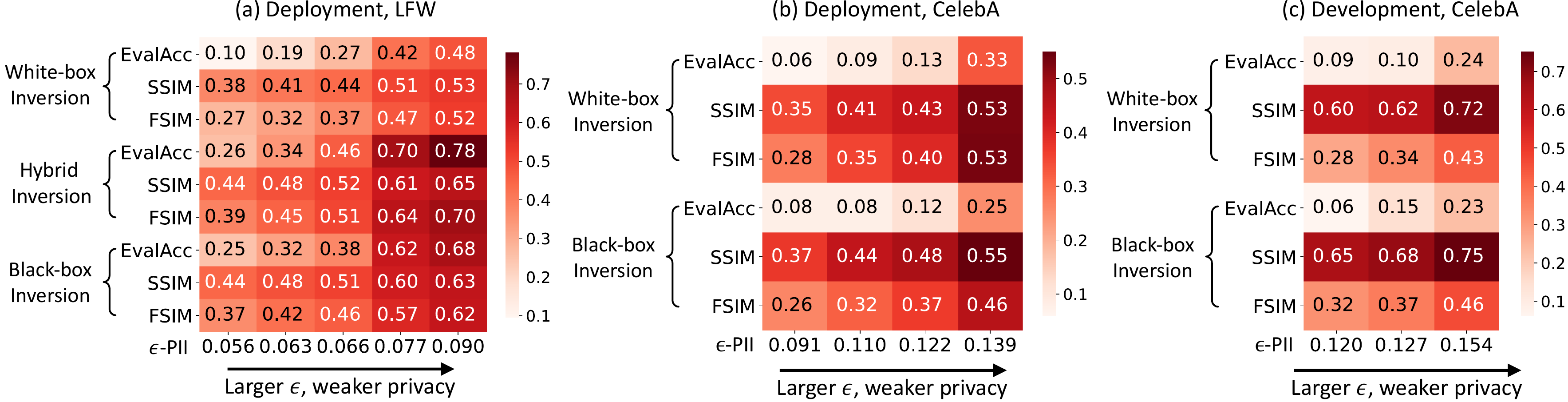}
    \caption{Crafter's PII on LFW and CelebA against different attacks, deployment and development scenario. A darker color indicates weaker empirical privacy.}
    \label{fig:emd}
    \end{figure*}

\noindent{\textbf{Validity of $\boldsymbol{\epsilon}$-PII.}}
We show that $\epsilon$-PII successfully reflects Crafter's empirical perceptual inversion privacy, supporting its validity as a perceptual privacy index for Crafter. The exact 
EMD between the  $G(z^*(F^*_X))$ distribution and the $G(z_r)$ distribution is intractable to compute,
so we compute its empirical approximation from the distribution samples with the \texttt{POT} solver \cite{pot}, and scale it down by the image size for simplicity. We show that this approximation error is bounded in Appendix \ref{sec:emd_appen} \cite{appendix}.
Figure~\ref{fig:emd} shows Crafter's result on LFW and CelebA under the model deployment and development scenario against different inversion attacks with different access.
As $\epsilon$ increases, all three empirical privacy metrics (Eval Acc, SSIM and FSIM) increase collaboratively regardless of the attacker instantiation, indicating weaker  empirical identity perceptual privacy against inversion attacks. 
\subsection{Defence against Adaptive Attacks}
\label{sec:adaptive}

\noindent\textbf{A1: Continue the optimization.}
We claim in \S\ref{sec:opt} that Adv Learning {  and Disco} do not provide any {worst-case guarantee of their simulated adversary} and is vulnerable against adaptive attacks, while Crafter successfully prevents adaptive attacks under the {  model deployment} scenario. Figure~\ref{fig:adaptive_white} and Figure~\ref{fig:adaptive_black} respectively show how Crafter defends against adaptive white-box and black-box attacks through iterations of update. 
{Basic attacks correspond to epoch 0 in each figure.}
In Figure~\ref{fig:adaptive_black}, line plots of Adv Learning starts at an Eval Acc and SSIM much lower than Crafter, indicating stronger privacy protection against basic attacks. As the black-box $Dec$ proceeds to update itself on $\mathcal{X}_{\mathrm{pub}}$, the privacy loss metrics drastically increases and ends up much higher than Crafter. 
{  Disco exhibits a similar performance: on LFW, its Eval Acc increases from 0 to 0.47, 0.60 and 0.63 for tradeoff parameter $\lambda=0.8,0.6$ and $0.2$ respectively. We omit the adaptive privacy evaluations of Disco on CelebA due to its impractical utility.}
In contrast, for Crafter, the Eval Acc and SSIM drop to or maintain at the same level with the basic black-box attacks. 
Similarly for the {  model development scenario}, TIPRDC also fails to defend against A1 adaptive attacks. Figure~\ref{fig:adaptive_black_tip} shows that Eval Acc of TIPRDC increases more than 15\% in 70 epochs, {while that of Crafter decreases slightly from the basic attacks}.  The experimental results support our claim (\textbf{Q2}).

\begin{figure}[ht]
    \centering
    \includegraphics[width=3.5in]{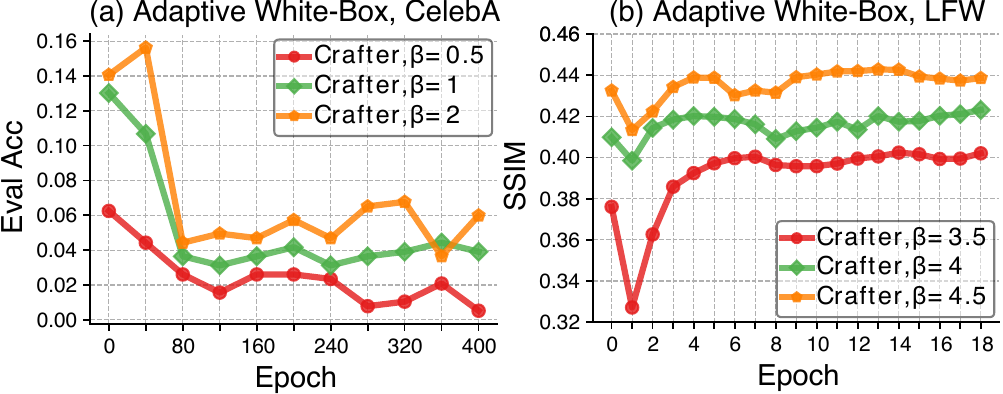}
    \caption{Crafter on CelebA and LFW against A1 adaptive white-box attacks, {deployment} scenario.}
    \label{fig:adaptive_white}
    \end{figure}
\begin{figure}[ht]
    \centering
    \includegraphics[width=3.5in]{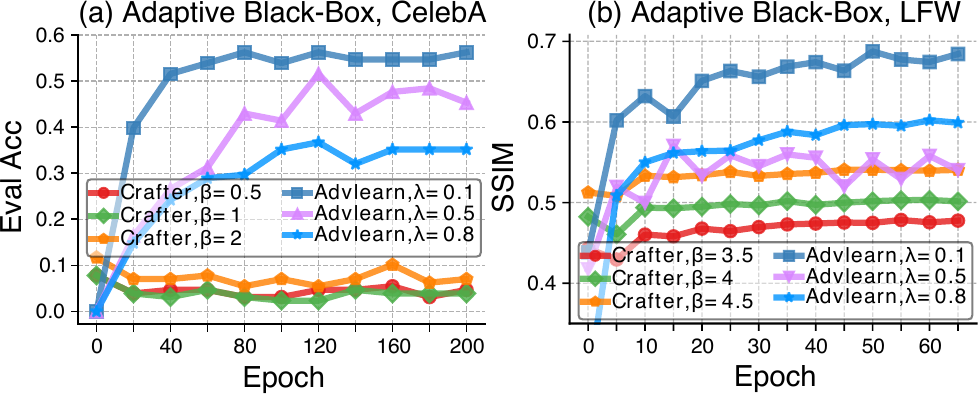}
    \caption{Crafter and Adv Learning on CelebA and LFW against A1 adaptive black-box attacks, {deployment} scenario.}
    \label{fig:adaptive_black}
    \end{figure}

We summarize how different schemes defend against attacks of varied strength in Figure~\ref{fig:heatmap}  and the corresponding visualization is in Figure~\ref{fig:heatvis} with more in Appendix~\ref{sec:vis} \cite{appendix}. Figure~\ref{fig:heatmap}(a) reports the average Eval Acc of the raw feature, Adv Learning and Crafter against five attacks on LFW {in the {deployment} scene.}
It is averaged across $\beta \in \{3.5, 4, 4.5\}$ for Crafter and $\lambda \in \{0.1,0.5,0.8\}$ for Adv Learning. Crafter exhibits robust privacy performance against all attacks, while Adv Learning is robust only against basic black-box and hybrid attacks. Figure~\ref{fig:heatmap}(b) draws a similar conclusion in comparison with TIPRDC in the training scene.
\begin{figure}[ht]
	\centering
	\includegraphics[width=3.5in]{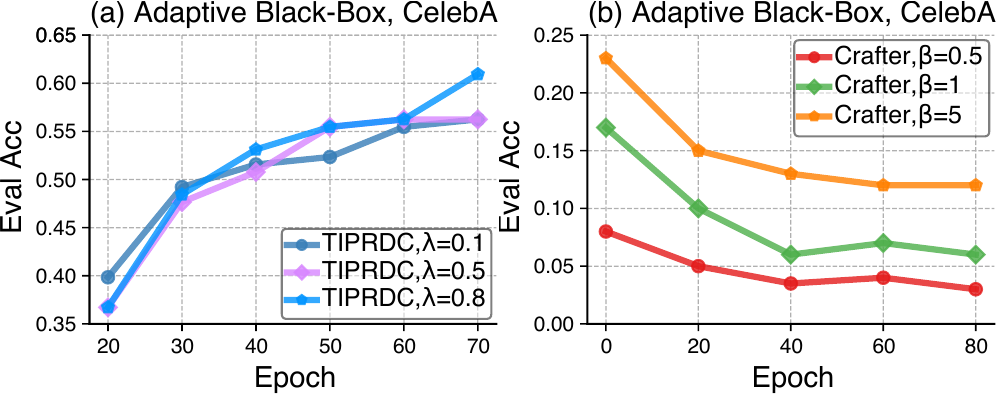}
	\caption{TIPRDC and {Crafter} on CelebA against black-box A1 adaptive attacks, {development} scenario.}
	\label{fig:adaptive_black_tip}     
\end{figure}

\begin{figure}[ht]
    \centering
	\includegraphics[width=\linewidth]{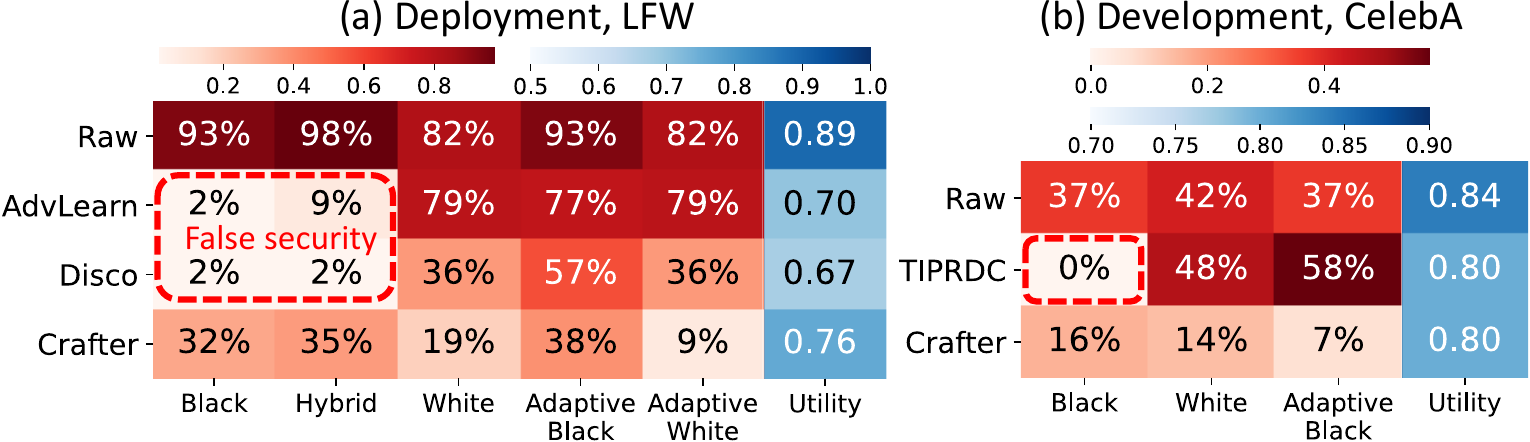}
    \caption{Average Eval Acc and utility AUC of defences on LFW and CelebA against different attacks, { deployment and development scenario. A darker red color indicates less robustness against attacks, and a darker blue indicates better utility}. {The dashed line indicates false security.}}     
    \label{fig:heatmap}
\end{figure}

As pointed out by Tramer et. al~\cite{tramer2020adaptive}, building a non-robust defence that prevents a particular attack is of little value (the ``no-free-lunch-theorem"). {The final Eval Acc that evaluates the protection strength should be the maximal value among all basic and adaptive attacks under all adversary access, so Crafter is superior to the baselines, as shown in Figure~\ref{fig:heatmap}.}

\noindent\textbf{A2: Utilize different generators.}
We show Crafter remains effective across attack models with different structures and $z$ dimensions. Specifically, Crafter uses generator $G_1$ and $z$ dimension of 500, whereas the adaptive attackers employ generators $G_1$ and $G_2$ across 5 different latent dimensions in the attacks. Figure~\ref{fig:gablation} shows that Crafter's Eval Acc on CelebA at $\beta = 1$ against different attacker models fluctuates within a threshold of 3\%, and FSIM differs no more than 0.01. We also evaluate Crafter against an adaptive attack with a more advanced model StyleGAN. Under Crafter protection, the attack achieves 0 success rate on CelebA, and does not beat basic attack on VGGFace2 either (Table~\ref{tab:vgg}). The `mean' entry reports the average SSIM of reconstruced images across the dataset, while the `worst' entry reports the largest SSIM. Hence 
Crafter is effective against A2 adaptive attacks (\textbf{Q2}).

\begin{table}[h]
	\small
	\centering
	\caption{SSIM of Crafter on VGGFace2 {against basic (WGAN) attack and A2 adaptive attack (StyleGAN)}.}
	\label{tab:vgg}
	\begin{tabular}{c||cc|cc||c}
	\hline
	$\beta$ & \multicolumn{2}{c|}{{Basic attack (WGAN)}} & \multicolumn{2}{c||}{A2 attack (StyleGAN)} & Utility \\ \hline
		      & mean        & worst        & mean          & worst         &         \\ \hline
	2         & 0.27        & 0.46         & 0.2           & 0.3           & 0.5     \\
	5         & 0.29        & 0.5          & 0.22          & 0.28          & 0.59    \\
	7         & 0.32        & 0.5          & 0.27          & 0.38          & 0.6875  \\ \hline
	\end{tabular}
	\end{table}

	\noindent\textbf{A3: Average features over multiple queries.}
We simulate an A3 adaptive attack that queries Crafter 5 times with the same batch of 64 images, and computes their mean as an averaged feature. The drastic increase from `Basic' to `Averaging' in Table~\ref{tab:average} shows that without shuffling, crafted perturbations on each query indeed offset each other weakening the defence. However, with a simple shuffling within the batch, the attack success rate is reduced to almost 0\%. { Therefore, to ensure robustness against A3 adaptive attack,
 we implement input shuffling as an inferent part of Crafter. 
 We acknowledge that passing A1, A2 and A3 does not guarantee Crafter bullet-proof. It is possible that more advanced attacks in the future  may corrupt the current defence.}


\begin{table}[]
	\centering
	\small
	\caption{Eval Acc of Crafter on CelebA against A3 adaptive attack. Batch shuffling is an essential user requirement to defend against averaging.}
	\label{tab:average}
	\resizebox{\columnwidth}{!}{
	\begin{tabular}{c|ccc}
		\hline
		$\beta$                     & 0.5  & 1     & 2         \\ \hline
		Basic {attack}& 5.98\% & 8.59\%  & 13.02\%  \\ \hline
		{A3 attack (Averaging)} & 10.16\% & 20.31\% & 26.56\%  \\\hline
		\makecell{{A3 attack against shuffling defence}} & 0\% & 0\% & 0.52\% \\\hline
	\end{tabular} }
\end{table}
\subsection{{Image-manipulation} Defences}
\label{sec:LowKey}
The image-manipulation defences are not strictly comparable with feature-manipulation schemes, but LowKey~\cite{cherepanova2021lowkey} and Fawkes~\cite{shan2020fawkes} share the same goal of private information concealing with Crafter. Hence we adopt the two schemes on private test sets to { evade identity classification models}. Specifically, poisoned images by their schemes are fed to $Enc$ and the produced features are being attacked.
{Figure \ref{fig:lowkey_visualize} shows that} the encoding and reconstruction procedure strips away Fawkes perturbation on images, leading to privacy loss metrics as poor as the raw feature. LowKey on LFW attains a comparable Eval Acc with a higher AUC than Crafter in Figure~\ref{fig:lowkey_visualize}. However, upon a closer look, the SSIM of reconstructed LowKey images are still as high as unprotected ones, meaning the attacker is able to reconstruct an image with high confidence although a facial verification model fails to predict its true identity. Evani et. al~\cite{carlinipoisoning} shows that even such an advantage over verification models can be overcome through robust training. The visualization (Figure \ref{fig:lowkey_images}) also supports the conclusion that such image-manipulation protection is ineffective against reconstruction attack. 
The black-box attack shows similar results.

As an alternative, the poisoned images can be generated to evade $Enc$. It achieves high privacy but disrupts data utility: the AUC drops to 0.52.
This is because drawing the user's protected feature close to that of another independent individual (Fawkes) or far away from the original feature (LowKey) causes large feature deviations, thus not preserving utility however small the image perturbation is.
Therefore, existing image-manipulation defences fail under our edge-cloud computing scenario.
\begin{figure}[h]
    \centering
    \begin{minipage}[t]{1.6in}\vspace{0pt}%
      \centering
      \includegraphics[width=1.6in]{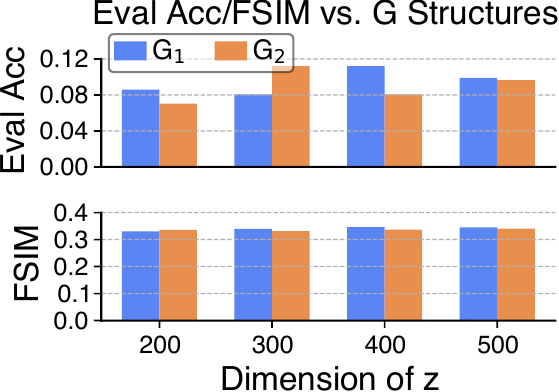}
      \caption{Crafter on CelebA against { basic attack ($G_1$, $z_{dim}=500$) and A2 adaptive attacks ($G_1, z_{dim}\in \{200,300,400\}$) and ($G_2$)}.}
      \label{fig:gablation}
    \end{minipage}
    \quad
    \begin{minipage}[t]{1.6in}\vspace{0pt}%
      \centering
      \includegraphics[width=1.6in]{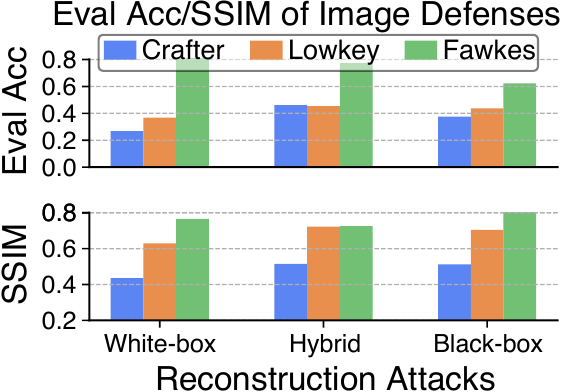}
      \caption{LowKey, Fawkes and Crafter on LFW against white-box, hybrid and black-box reconstruction attacks.}
      \label{fig:lowkey_visualize} 
    \end{minipage}
  \end{figure}

  \begin{figure}[ht]
    \centering
    \includegraphics[width=0.9\linewidth]{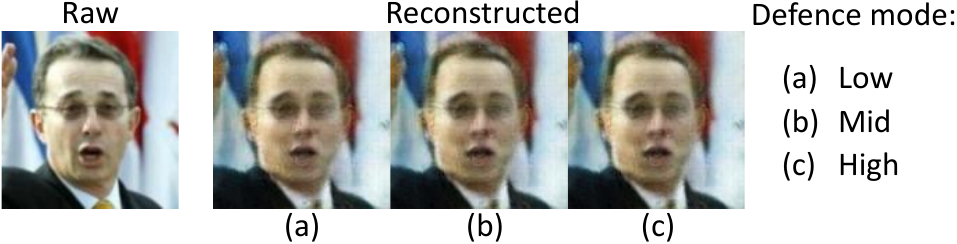}
    \caption{Reconstructed images under  LowKey protection. LowKey fails to preserve pixel-level privacy regardless of the defence mode.}
    \label{fig:lowkey_images}
    \end{figure}  
\subsection{User study and Running Time} 
\noindent\textbf{Human study.}
Besides using the evaluating network as an oracle, we conduct a human study to further evaluate if the inverted images under attacks can be recognized by human.
A sample question is in 
Figure \ref{fig:human}, where participants are given one reconstructed LFW image under Crafter’s protection with $\beta=4.5$ 
(marked with an asterisk). 
At most one of the options belongs to the same identity as the protected image, and participants may choose “None above” if they believe options A-E do not cover the correct answer.
Hence each option has an equivalent rate of 20\% to be correct. Please
refer to Appendix \ref{appen:human} \cite{appendix} for more details of the study protocol.
\begin{figure}[ht]
	\centering
	\includegraphics[width=0.95\linewidth]{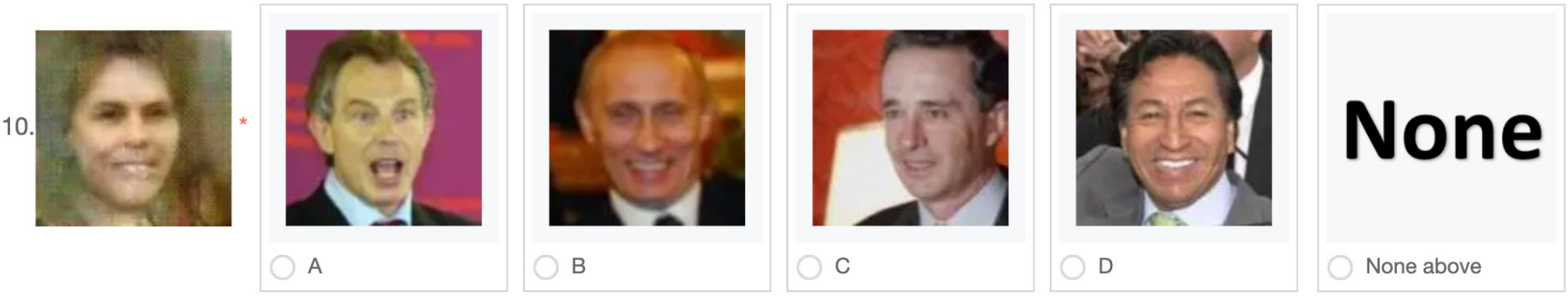}
	\caption{A sample question from the human study poll.}
	\label{fig:human}
\end{figure}

We eventually harvest 35 valid feedbacks each from a different individual as shown in Table \ref{tab:human1}.
The Macro-F1 measure is 0.251, which is close to that of a random guess (0.200). Thus Crafter is effective by human evaluation. 
\begin{table}[ht]
    \small
    \centering
    \caption{Human study results.}
    \begin{tabular}{c|c|ccccc}
    \hline
    \multicolumn{2}{c|}{\multirow{2}{*}{35 participants}} & \multicolumn{5}{c}{Participant's choice}     \\ \cline{3-7}
    \multicolumn{2}{c|}{}        & A  & B & C& D&`None'      \\
	\hline
    \multirow{5}{*}{Ground Truth}               & A   & 7    & 11    & 11 & 17&24 \\\cline{2-7}
    \multirow{5}{*}{}                 & B   & 17   & 8    & 21&6&18 \\\cline{2-7}
    \multirow{5}{*}{}               & C   &15&4&25&11&15  \\\cline{2-7}
    \multirow{5}{*}{}                & D   &7 &10&4&24&25 \\\cline{2-7}
    \multirow{5}{*}{}                  & `None'  &15&9&11&8&27 \\ \hline
    \end{tabular}
    \label{tab:human1}
\end{table}

\noindent\textbf{Running time.}
Crafter runs upon every incoming batch, and it is unrealistic to benchmark it against schemes involving network training (\emph{e.g.,} Adv Learning, TIPRDC). Hence, we compare Crafter's running time with a similar setting in LowKey, where the 128$\times$128 LFW image is crafted. The results are shown in Table~\ref{tab:runtime} that Crafter's running time is comparable with existing ones (\textbf{Q4}). In Crafter, the feature crafting iteration no. is 500 and the {\tt approxInverseHVP} iteration no. is 150. We observe the runtime bottleneck is the inversion and {\tt approxInverseHVP} steps (lines 3 and 11 in Alg.~\ref{alg1}). Thus we speedup the inversion by training an amortizer offline (see details in Appendix~\ref{sec:amor} \cite{appendix}). Improving the speed of the latter is left to future work, which now takes up 71.5\% of the total running time.

\begin{table}[]
	\centering
    \small
    \caption{Running time comparison between Crafter and Lowkey with varied feature/image sizes, on LFW.}
    \label{tab:runtime}
	\begin{tabular}{c|c|c}
		\hline
		Defense                  & Feature (Image) Size  & Runtime(s)/Image \\ \hline
		{Crafter}    & (64, 16, 16)  & 27.15             \\ \cline{2-3} 
		& (128, 16, 16) & 30.82             \\ \cline{2-3} 
		& (64, 32, 32)  & 53.71             \\ \cline{1-3} 
		\multicolumn{1}{l|}{LowKey} & (3, 128, 128)             & 87.45             \\ \hline
	\end{tabular}
\end{table}

\renewcommand\tabcolsep{3.0pt}
\begin{table}[]
	\centering
    \small
    \caption{Eval Acc of Crafter against white-box attack w/wo private image exposure, and w/wo the original feature exposure, on CelebA.}
    \label{tab:expose}
	\begin{tabular}{c|ccc}
		\hline
		$\beta$                     & 0.5  & 1     & 2         \\ \hline
		Crafter & 5.98\% & 8.59\%  & 13.02\%  \\ \hline
		With $\mathcal{X}_{\mathrm{pvt}}$ & 6.77\% & 10.93\% & 13.28\% \\\hline
		With $Enc(X)$ & 9.38\% & 17.97\% & 21.09\% \\\hline
	\end{tabular} 
\end{table}

\subsection{Limitations}
\label{sec:extended}
We discuss Crafter's limitations to further specify its usage.

\noindent\textbf{Private image exposure.} Ideally, private images are not publicly available, but we discuss how the accidental exposure of private images would affect Crafter. We expose 10\% of each individual's private images to the white-box attacker and retrain its generator to simulate a realistic exposure. Comparions between entry `Crafter' and `With $\mathcal{X}_{\mathrm{pvt}}$' in Table~\ref{tab:expose} shows the Eval Acc of Crafter on CelebA against white-box attacks with and without private image exposure. There is a slight increase of 2\% on Eval Acc, but the impact is minor overall. 

\noindent\textbf{Exposure of original features.}
It is also a threat if the attacker happens to intercept the original features $Enc(X)$ of a set of private images, based on which it trains a feature-to-identity classification network, and predict the ID of private $F_X$ directly. The two groups of private images are different but having identity overlaps. We simulate the case by letting the attacker collect $(Enc(\mathcal{X}_{\mathrm{train}}), ID)$ pairs to train an ID classifier, and infer the ID by feeding in $F_{\mathcal{X}_{\mathrm{test}}}$. The evaluating accuracy is given in Table~\ref{tab:expose}, which is inferior to Crafter's original performance. Hence it is important not to reveal the corresponding identity label of private features $Enc(X)$ for the effectiveness of Crafter. We consider this requirement attainable, as users are not motivated to share the private identity label anywhere in our problem setting.

\section{Related Works}
\label{sec:relwork}
We focus on prior works preserving input privacy in deep neural networks (DNNs), especially when inputs are images.

\subsection{Image Perturbation.}
To prevent the identity of an image from being disclosed in DNN processing, de-identification is proposed to alter the raw image. One technique to achieve de-identification is adversarial image perturbation, where visually insignificant perturbation is crafted to disrupt the prediction result of an ensemble of identity classifiers~\cite{cherepanova2021lowkey,shan2020fawkes}. Although these approaches are effective against state-of-the-art auto-recognition models such as {Microsoft Azure Face API~\cite{azure} and Amazon Rekognition~\cite{amazon},} they fail to preserve visual privacy, i.e. the perturbed image is close to the original one and any semantic information could be exposed. {Zheng \emph{et al.}~\cite{cispa} prevents GAN-inversion-based facial image manipulation with imperceptible image perturbation that maximizes the distance between the original and pertubed images in the latent and feature space. This is a scenario opposite to that of Crafter:~\cite{cispa} pushes the protected image away from the original in feature space to protect privacy, while Crafter draws the feature close to the original to preserve utility.} 
Wu \emph{et al.}~\cite{wu2021dapter} protects visual privacy by a transformer network \emph{DAPter} that generates images with low image entropy while minimizing inference loss to preserve image utility. However, it is restricted to specified learning tasks and does not defend inversion attacks.
 Different from image perturbation, we choose to preserve privacy by injecting perturbations to features.
\subsection{Feature Perturbation.}
An alternative to image perturbation/obfuscation is to send an encoded feature of the corresponding input to serve the downstream tasks instead of raw images. The feature, on one hand, carries much information of the input images, on the other, prevents direct revealing of raw inputs. To prevent adversaries from reconstructing inputs or inferring private attributes from features, several works propose to simulate a game between the attacker and protector with conflicting privacy goals: the protector fights against the worst-case attacker by producing features which the attacker would fail to invert. Li \emph{et al.}~\cite{li2021deepobf}, Xiao \emph{et al.}~\cite{xiao2020adversarial}, {  Singh \emph{et al.}~\cite{singh2021disco}} and Wang \emph{et al.}~\cite{wang2021MI} propose to learn perturbed features with adversarial networks and only upload those features for downstream DNN tasks. They either minimize the resemblance of reconstructed images to original ones~\cite{xiao2020adversarial,li2021deepobf}, or minimize the mutual information between the obfuscated feature and the raw input~\cite{wang2021MI}. While sharing a similar idea of preserving utility by minimizing inference accuracy loss, the aforementioned works all suffer the drawback of requiring specified inference tasks in the adversarial training. Being task-independent, TIPRDC~\cite{li2020tiprdc} {  and Decouple~\cite{ragan2011decoupled} }lift the constraint by maximizing the mutual information between features and raw inputs to preserve utility, and thus require no knowledge of the learning task. However, they demand the private information be specifically defined with labels, {i.e.,} a man with or without a beard. Our method is designed to defend against identity theft, in which the private attribute is challenging to isolate from the raw input.

The above adversarial-networks-based solutions share a common defect: there is no guarantee for the convergence point of the adversarial game, rendering them vulnerable to adaptive attackers in \S\ref{sec:motivation}. In addition, they all involve retraining the backend model on the cloud, which is typically costly. In comparison, our framework is robust against adaptive attackers, and requires no change in the cloud backend.

\section{Conclusion}
\label{sec:conclusion}
We present Crafter, a facial feature crafting system against inversion attacks in deep learning models. We define privacy as the distributional distances between the attacker's posterior and prior views on the input facial images given the feature. As feature is implicitly expressed in the privacy-utility joint optimization objective, we take an IFT-based approach to solve the problem. Analysis and experimental results support that Crafter successfully defends against a variety of attacks with little computation accuracy loss.

\section*{Acknowledgment}
The authors would like to thank the reviewers for their constructive comments. This work was supported in part by National Key R\&D Program of China under Grant 2021ZD0112801, NSF China (62272306, 62032020, 62136006), a RGC RIF grant under the contract R6021-20, RGC CRF grants under the contracts C7004-22G and C1029-22G, and RGC GRF grants under the contracts 16209120, 16200221 and 16207922. Authors would like to appreciate the Student Innovation Center of SJTU for providing GPUs. Shiming Wang personally thanks Ruiming Lu for feedbacks on early versions of this paper.

\bibliographystyle{IEEEtranS}
\bibliography{IEEEabrv}

\begin{thebibliography}{10}
\providecommand{\url}[1]{#1}
\csname url@samestyle\endcsname
\providecommand{\newblock}{\relax}
\providecommand{\bibinfo}[2]{#2}
\providecommand{\BIBentrySTDinterwordspacing}{\spaceskip=0pt\relax}
\providecommand{\BIBentryALTinterwordstretchfactor}{4}
\providecommand{\BIBentryALTinterwordspacing}{\spaceskip=\fontdimen2\font plus
\BIBentryALTinterwordstretchfactor\fontdimen3\font minus
  \fontdimen4\font\relax}
\providecommand{\BIBforeignlanguage}[2]{{%
\expandafter\ifx\csname l@#1\endcsname\relax
\typeout{** WARNING: IEEEtranS.bst: No hyphenation pattern has been}%
\typeout{** loaded for the language `#1'. Using the pattern for}%
\typeout{** the default language instead.}%
\else
\language=\csname l@#1\endcsname
\fi
#2}}
\providecommand{\BIBdecl}{\relax}
\BIBdecl

\bibitem{amazon}
``Amazon rekognition face verification api.''
  \url{https://aws.amazon.com/rekognition/}.

\bibitem{azure}
``Microsoft azure face api.'' \url{https://azure.microsoft.com/en-us/services/
  cognitive-services/face/}.

\bibitem{appendix}
``Our crafter.''
  \url{https://github.com/ShimingWang98/Facial_Feature_Crafting_against_Inversion_based_Identity_Theft/tree/main}.

\bibitem{abadi2016deep}
M.~Abadi, A.~Chu, I.~Goodfellow, H.~B. McMahan, I.~Mironov, K.~Talwar, and
  L.~Zhang, ``Deep learning with differential privacy,'' in \emph{Proc.of ACM
  SIGSAC}, 2016.

\bibitem{key}
S.~An, G.~Tao, Q.~Xu, Y.~Liu, G.~Shen, Y.~Yao, J.~Xu, and X.~Zhang, ``Mirror:
  Model inversion for deep learning network with high fidelity,'' in
  \emph{Proc. of NDSS}, 2022.

\bibitem{arjovsky2017wasserstein}
M.~Arjovsky, S.~Chintala, and L.~Bottou, ``Wasserstein generative adversarial
  networks,'' in \emph{Proc.of ICML}.

\bibitem{blau2018perception}
Y.~Blau and T.~Michaeli, ``The perception-distortion tradeoff,'' in
  \emph{Proceedings of the IEEE conference on computer vision and pattern
  recognition}, 2018, pp. 6228--6237.

\bibitem{rdp}
------, ``Rethinking lossy compression: The rate-distortion-perception
  tradeoff,'' in \emph{International Conference on Machine Learning}.\hskip 1em
  plus 0.5em minus 0.4em\relax PMLR, 2019, pp. 675--685.

\bibitem{carlini2021privateencoding}
N.~Carlini, S.~Deng, S.~Garg, S.~Jha, S.~Mahloujifar, M.~Mahmoody, A.~Thakurta,
  and F.~Tram{\`e}r, ``Is private learning possible with instance encoding?''
  in \emph{Proc.of 2021 IEEE S\&P}, 2021.

\bibitem{chen2021perceptual}
J.~Chen, L.~Chen, C.~Yu, and C.~Lu, ``Perceptual indistinguishability-net
  (pi-net): Facial image obfuscation with manipulable semantics,'' in
  \emph{Proc.of CVPR}, 2021.

\bibitem{cherepanova2021lowkey}
V.~Cherepanova, M.~Goldblum, H.~Foley, S.~Duan, J.~Dickerson, G.~Taylor, and
  T.~Goldstein, ``Lowkey: leveraging adversarial attacks to protect social
  media users from facial recognition,'' in \emph{Proc.of ICLR}, 2021.

\bibitem{dusmanu2020sift}
M.~Dusmanu, J.~L. Sch{\"{o}}nberger, S.~N. Sinha, and M.~Pollefeys,
  ``Privacy-preserving image features via adversarial affine subspace
  embeddings.'' in \emph{Proc.of CVPR}, 2020.

\bibitem{pot}
R.~Flamary, N.~Courty, A.~Gramfort, M.~Z. Alaya, A.~Boisbunon, S.~Chambon,
  L.~Chapel, A.~Corenflos, K.~Fatras, N.~Fournier, L.~Gautheron, N.~T. Gayraud,
  H.~Janati, A.~Rakotomamonjy, I.~Redko, A.~Rolet, A.~Schutz, V.~Seguy, D.~J.
  Sutherland, R.~Tavenard, A.~Tong, and T.~Vayer, ``Pot: Python optimal
  transport,'' in \emph{Journal of Machine Learning Research}, vol.~22, no.~78,
  2021, pp. 1--8.

\bibitem{fredrikson2015model}
M.~Fredrikson, S.~Jha, and T.~Ristenpart, ``Model inversion attacks that
  exploit confidence information and basic countermeasures,'' in \emph{Proc. of
  ACM SIGSAC}, 2015.

\bibitem{Gulrajani2017wgangp}
I.~Gulrajani, F.~Ahmed, M.~Arjovsky, V.~Dumoulin, and A.~C. Courville,
  ``Improved training of wasserstein gans,'' in \emph{Proc.of NIPS}, 2017.

\bibitem{huang2020instahide}
Y.~Huang, Z.~Song, K.~Li, and S.~Arora, ``Instahide: Instance-hiding schemes
  for private distributed learning,'' in \emph{Proc.of ICML}, 2020.

\bibitem{li2020tiprdc}
A.~Li, Y.~Duan, H.~Yang, Y.~Chen, and J.~Yang, ``Tiprdc: task-independent
  privacy-respecting data crowdsourcing framework for deep learning with
  anonymized intermediate representations,'' in \emph{Proc.of ACM SIGKDD},
  2020.

\bibitem{li2021deepobf}
A.~Li, J.~Guo, H.~Yang, F.~D. Salim, and Y.~Chen, ``Deepobfuscator: Obfuscating
  intermediate representations with privacy-preserving adversarial learning on
  smartphones,'' in \emph{Proc.of IOTDI}, 2021.

\bibitem{liu2015LFW}
Z.~Liu, P.~Luo, X.~Wang, and X.~Tang, ``Deep learning face attributes in the
  wild,'' in \emph{Proc.of CVPR}, 2015.

\bibitem{lorraine2020optimizing}
J.~Lorraine, P.~Vicol, and D.~Duvenaud, ``Optimizing millions of
  hyperparameters by implicit differentiation,'' in \emph{Proc.of AISTATS},
  2020.

\bibitem{lowe2004distinctive}
D.~G. Lowe, ``Distinctive image features from scale-invariant keypoints,'' in
  \emph{Proc.of INT J COMPUT VISION}, 2004.

\bibitem{mahendran2014understanding}
A.~Mahendran and A.~Vedaldi, ``Understanding deep image representations by
  inverting them,'' in \emph{Proc.of cs.CV}, 2014.

\bibitem{black_inv}
F.~M. Orekondy~T, Schiele~B, ``Knockoff nets: Stealing functionality of
  black-box models,'' in \emph{Proc. of CVPR}, 2019.

\bibitem{carlinipoisoning}
T.~F. Radiya-Dixit~E, Sanghyun~Hong, ``Data poisoning won't save you from
  facial recognition,'' 2022.

\bibitem{ragan2011decoupled}
J.~Ragan-Kelley, J.~Lehtinen, J.~Chen, M.~Doggett, and F.~Durand, ``Decoupled
  sampling for graphics pipelines,'' in \emph{ACM Transactions on Graphics
  (TOG)}, vol.~30, no.~3.\hskip 1em plus 0.5em minus 0.4em\relax ACM New York,
  NY, USA, 2011, pp. 1--17.

\bibitem{schroff2015facenet}
F.~Schroff, D.~Kalenichenko, and J.~Philbin, ``Facenet: A unified embedding for
  face recognition and clustering,'' in \emph{Proc.of CVPR}, 2015.

\bibitem{ali2018poison}
A.~Shafahi, W.~R. Huang, M.~Najibi, O.~Suciu, C.~Studer, T.~Dumitras, and
  T.~Goldstein, ``Poison frogs! targeted clean-label poisoning attacks on
  neural networks,'' in \emph{Proc.of NIPS}, 2018.

\bibitem{shan2020fawkes}
S.~Shan, E.~Wenger, J.~Zhang, H.~Li, H.~Zheng, and B.~Y. Zhao, ``Fawkes:
  Protecting privacy against unauthorized deep learning models,'' in
  \emph{Proc.of USENIX Security}, 2020.

\bibitem{singh2021disco}
A.~Singh, A.~Chopra, E.~Garza, E.~Zhang, P.~Vepakomma, V.~Sharma, and
  R.~Raskar, ``Disco: Dynamic and invariant sensitive channel obfuscation for
  deep neural networks,'' in \emph{Proceedings of the IEEE/CVF Conference on
  Computer Vision and Pattern Recognition}, 2021, pp. 12\,125--12\,135.

\bibitem{tramer2020adaptive}
B.~W. Tramer~F, Carlini~N, ``On adaptive attacks to adversarial example
  defenses,'' in \emph{Proc.of NIPS}, 2020.

\bibitem{wang2021datalens}
B.~Wang, F.~Wu, Y.~Long, L.~Rimanic, C.~Zhang, and B.~Li, ``Datalens: Scalable
  privacy preserving training via gradient compression and aggregation,'' in
  \emph{Proc.of ACM SIGSAC}, 2021.

\bibitem{perceptual}
T.~Wang, Y.~Zhang, Y.~Fan, J.~Wang, and Q.~Chen, ``High-fidelity gan inversion
  for image attribute editing,'' in \emph{Proceedings of the IEEE/CVF
  Conference on Computer Vision and Pattern Recognition (CVPR)}, June 2022, pp.
  11\,379--11\,388.

\bibitem{wang2021MI}
T.~Wang, Y.~Zhang, and R.~Jia, ``Improving robustness to model inversion
  attacks via mutual information regularization.'' in \emph{Proc.of AAAI},
  2020.

\bibitem{wang2016celeba}
Z.~Wang, S.~Chang, Y.~Yang, D.~Liu, and T.~S. Huang, ``Studying very low
  resolution recognition using deep networks,'' in \emph{Proc.of CVPR}, 2016.

\bibitem{wei2020federated}
K.~Wei, J.~Li, M.~Ding, C.~Ma, H.~H. Yang, F.~Farokhi, S.~Jin, T.~Q. Quek, and
  H.~V. Poor, ``Federated learning with differential privacy: Algorithms and
  performance analysis,'' in \emph{IEEE Transactions on Information Forensics
  and Security}, 2020.

\bibitem{wu2021dapter}
H.~Wu, X.~Tian, Y.~Gong, X.~Su, M.~Li, and F.~Xu, ``Dapter: Preventing user
  data abuse in deep learning inferenceservices,'' in \emph{Proc.of WWW}, 2021.

\bibitem{xiao2020adversarial}
T.~Xiao, Y.-H. Tsai, K.~Sohn, M.~Chandraker, and M.-H. Yang, ``Adversarial
  learning of privacy-preserving and task-oriented representations,'' in
  \emph{Proc.of AAAI}, 2020.

\bibitem{zhang2020secret}
Y.~Zhang, R.~Jia, H.~Pei, W.~Wang, B.~Li, and D.~Song, ``The secret revealer:
  Generative model-inversion attacks against deep neural networks,'' in
  \emph{Proc.of cs.LG}, 2020.

\bibitem{cispa}
L.~Zheng, Y.~Ning, A.~Salem, M.~Backes, M.~Fritz, and Z.~Yang, ``Unganable:
  Defending against gan-based face manipulation,'' in \emph{Proc.of USENIX
  Security}, 2023.

\end{thebibliography}

\appendices
\newpage
\section{Threat Model Clarifications and Other Related Works}
\label{sec:rw}
\subsection{Clarifications on the Threat Model}
\label{sec:appen_comparison}

Three of the most prevailing attacks against user privacy is membership inference attack, attribute inference attack and model inversion attack. 
 We compare Crafter with some most representative baselines in Table \ref{tab:comparison_realted} to further clarify the threat model in our scenario.

 Crafter and Adv Learning~\cite{xiao2020adversarial} specifically defend against model inversion attackers that are interested in the victim user's unknown identity. The attacker only has access to some public images that are non-overlapping with the user's identity, and intend to unveil the user's appearance via reconstruction. 
 
 Disco~\cite{singh2021disco} and TIPRDC~\cite{li2020tiprdc} defend both model inversion and attribute inference attacks that aim to attain the users private attributes (eg. age and gender). 
  Specifically, the attribute inference attacker trains an attribute classifer with an available training set. We change the sensitive attribute to `identity' in our evaluation against inversion attacks.

  Fawkes~\cite{shan2020fawkes} and LowKey~\cite{cherepanova2021lowkey} also defend against attribute inference attacks, but the private attribute of interest is identity.
  However, their identity attribute privacy is significantly different from our identity privacy against inversion attack. In LowKey and Fawkes, the attacker already has access to some of the victim's private images, on which it trains an attribute inference (facial recognition) attacker model  to infer the identity of the victim's other images. The defence is considered effective as long as the facial recognition attacker fails. In contrast, Crafter's attacker is interested in the appearance of an unknown victim, and the defence is robust only if the reconstructed image reveals no identity information percetually. Hence in Crafter, the facial recognition model merely serves as an oracle to evaluate the perceptual similarity between reconstructed and the raw images, but cannot be a real-world adversary model as private IDs of the training data are missing.

    \begin{table}[ht]
        \centering
        \small
        \caption{Comparison between Crafter and existing defences.}
        \label{tab:comparison_realted}
        \resizebox{\columnwidth}{!}{
        \begin{tabular}{cccc}
            \toprule
            Defense & Threat model & Privacy & Utility \\
            \midrule
            Crafter & \multirow{2}{*}{Model inversion} & \multirow{2}{*}{Identity} & Task oriented/agnostic \\ 
            AdvLearn~\cite{xiao2020adversarial} & & & Task oriented \\
            \cmidrule{1-4}
            Disco~\cite{singh2021disco} & Model inversion \& & \multirow{2}{*}{Attribute} & Task oriented \\
            TIPRDC~\cite{li2020tiprdc} & Attribute inference & & Task agnostic \\
            \cmidrule{1-4}
            Fawkes~\cite{shan2020fawkes} & Attribute inference & {Attribute} & \multirow{2}{*}{/}\\
            LowKey~\cite{cherepanova2021lowkey} & (Attribute=Identity) & (Identity)&  \\
            \bottomrule
        \end{tabular}
        }
    \end{table}

{Besides those discussed in \S\ref{sec:relwork}, our work is also related to the following works.}
\subsection{Image Obfuscation}
Another line of work protects users' private information by obfuscating their raw images. 
\emph{InstaHide}~\cite{huang2020instahide} successfully preserves the visual privacy of raw images by mixing multiple images and randomly flipping the signs of the pixels. However, an advanced reconstruction attack~\cite{carlini2021privateencoding} can recover nearly visually identical reconstructions to the private images from the obfuscated ones by \emph{InstaHide}. Other work~\cite{wang2021datalens,wei2020federated,abadi2016deep} adopt differential privacy (DP), which is a strong privacy guarantee for an individual's data against membership inference attacks and training data memorization. However, the DP guarantee is different from our identity privacy. For example, \emph{DataLens}~\cite{wang2021datalens} leverages DP to generate synthetic DNN training data from users' private images, such that two datasets differing by one image sample have close probabilities to generate the same output DNN. Compared to DP, our goal is to prevent the attacker from inferring protected identity information of an image in DNN processing.

\subsection{Feature Obfuscation}
Another line of works obfuscates the private feature with other non-private ones to confuse the attacker. Dusmanu \emph{et al.}~\cite{dusmanu2020sift} perform obfuscation by affine space embedding, and Chen \emph{et al.}~\cite{chen2021perceptual} extends the classic Differential Privacy mechanism to image features for obfuscation. However, the former~\cite{dusmanu2020sift} is designed for image description and visual positioning (e.g. SIFT descriptor~\cite{lowe2004distinctive}) rather than DNN processing, while the latter~\cite{chen2021perceptual} is confined by the requirement of specifying the utility attributes in prior and thus can not fulfill unknown downstream computation tasks.

\section{Details of White-box Attack}
\label{appen:white}
The adversary aims to find through optimization an image $\hat{X}^*$ that best resembles the input $X$ from the viewpoint of encoder representation~\cite{mahendran2014understanding}
, {\em i.e.,} $\mathop{\min}_{\hat{X}} \|Enc(\hat{X}) - Enc(X)\|_2$. 
Finding $\hat{X}^*$ from scratch via gradient descent is extremely ill-posed or easily ends up at some visually meaningless local minimum. Therefore, the attackers can exploit public datasets ({\em e.g.,} facial images crawled from the Internet that are irrelevant to the private input $X$) to extract prior knowledge about general facial images. Specifically, the white-box attacker distills the public prior using the canonical Wasserstein-GAN structure \cite{Gulrajani2017wgangp} as proposed by Zhang \cite{zhang2020secret},  where a generator $G: \mathbb{R}^d \to \mathbb{R}^{w\times h}$ and discriminator $D: \mathbb{R}^{w\times h} \to \mathbb{R}$ are pretrained on public dataset $\mathcal{X}_{\mathrm{pub}}$ that has no identity overlapping with users' private $\mathcal{X}_{\mathrm{pvt}}$:
\begin{equation}
    \label{eq:pretrain}
    \mathop{\min}_{G}  \mathop{\max}_{D} ~ \mathbb{E}_{X'\in \mathcal{X}_{\mathrm{pub}}}[D(X')] - \mathbb{E}_z[D\circ G(z)].
\end{equation}
 Initiated from this public prior, searching for the best-suited image $\hat{X}^*$ can be transformed into gradient-based optimization on latent representation $z${, starting from some random $z_0$}:
\begin{equation}
    z^*=\arg\mathop{\min}_{z} \|Enc(X) - Enc\circ G(z)\|_2,
\end{equation}
and the reconstructed image is 
$
    \hat{X}^* = G(z^*).
$

\section{Implicit Gradient Computation}
\label{appen:ift}
Here we present the detailed computation of the implicit gradient $\frac{\partial \mathcal{L}_{p}^{*}(D,z^*(F_X))}{\partial {{F_X}}}$.
For $\left({F_X'}, z^{\prime}\right)$ that satisfies 
    \begin {enumerate*} [1)]
    \item
    $
      \left.\frac{\partial \mathcal{L}_{\mathrm{inv}}(F_X, z)}{\partial z} \right|_ 
    {{F_{\rm{x}}'}, z^{\prime}}=0$
    and 
    \item the Jacobian matrix of $\frac{\partial \mathcal{L}_{\mathrm{inv}}}{\partial z}$, $i.e. \left[\frac{\partial^{2} \mathcal{L}_{\mathrm{inv}}}{\partial z \partial z}
    \right]$ 
    is invertible,
    \end {enumerate*}
     surrounding $(F_X', z')$ we can write $z^*(F_X)$ as a function of $F_X$  s.t. $\left.\frac{\partial \mathcal{L}_{\mathrm{inv}}}{\partial z}\right|_ 
    {{F_X}, z^*(F_X)}=0$ and
    \begin{equation}
        \label{eq:IFT}
        \left. \frac{\partial z^{*}}{\partial {F_X}}\right
        |_{{F_X'}}=
        -\left[\frac{\partial^{2} \mathcal{L}_{\mathrm{inv}}}{\partial z \partial z}
        \right]^{-1}
        \times\left.\frac{\partial^{2} \mathcal{L}_{\mathrm{inv}}}{\partial z \partial {F_X}}\right|
        _{{F_X}^{\prime}, z^{*}\left({F_X'}\right)}.
    \end{equation}
Condition 
\begin {enumerate*} [1) ]
    \item
    $
      \left.\frac{\partial \mathcal{L}_{\mathrm{inv}}}{\partial z}\right|_{{F_{\rm{x}}'}, z^{\prime}}
      =0
      $
\end{enumerate*}
can be satisfied as the first-order condition is met in optimizing $\mathcal{L}_{\mathrm{inv}}$. As for condition 2), exactly inverting the Hessian in Eq.~(\ref{eq:IFT}) introduces large computational overhead in high-dimensional space, and hence we adopt a tractable inverse Hessian approximation by Lemma~\ref{lemma:neumann}:
\begin{equation}
    \left[\frac{\partial^{2} \mathcal{L}_{\mathrm{inv}}}{\partial z \partial z}
    \right]^{-1}=
\alpha \lim _{i \rightarrow \infty} \sum_{j=0}^{i}\left[I-
\alpha\frac{\partial^{2} \mathcal{L}_{\mathrm{inv}}}{\partial z \partial z}\right]^{j}
\end{equation}
where $\alpha$ is sufficiently small such that 
$|I-
\alpha\frac{\partial^{2} \mathcal{L}_{\mathrm{inv}}}{\partial z \partial z}|<1$.
Putting them together, we have
\begin{align}
    &\frac{\partial \mathcal{L}_{p}^{*}(D,z^*(F_X))}{\partial {{F_X}}}=
    \frac{\mathcal{L}_{p}^{*}(D,{z^*(F_X)})}
    {\partial z^*(F_X)} 
    \frac{\partial z^*(F_X)}{\partial {F_X}} \\
    &=
    -\alpha \frac{\mathcal{L}_{p}^{*}(D,{z^*})}
    {\partial z^*} 
    \cdot
    \lim _{i \rightarrow \infty} \sum_{j=0}^{i}\left[I-
    \alpha\frac{\partial^{2} \mathcal{L}_{\mathrm{inv}}}{\partial z \partial z}\right]^{j}
    \cdot
    \frac{\partial^{2} \mathcal{L}_{\mathrm{inv}}}{\partial z \partial {F_X}}. \notag
\end{align}

\newcommand{\x}[0]{z_\text{avg}}
\newcommand{\px}[0]{p_\text{avg}}
\newcommand{\pxhat}[0]{\hat{p}_\text{avg}}
\newcommand{\vvv}[0]{F_X}
\newcommand{\vstar}[0]{F_X^*}\newcommand{\z}[0]{z_0}
\newcommand{\dnn}[0]{d_{nn}}

\section{Privacy Guarantee}
\subsection{{The PII Guarantee of Crafter}} \label{appen:proof}
We denote the discriminator class as ${\mathcal{F}_D}$, and use $\mathcal{H}_{F_X\times D}$ to represent the compositional function class of inverting a feature $F_X$ into an image and feeding it into a discriminator.
Then the distance between Crafter's PII and the optimal attainable PII is  bounded by the Rademacher complexity $\mathcal{R}$ of  ${\mathcal{F}_D}$ and  $\mathcal{H}_{F_X\times D}$ as below.

\begin{theorem}[{PII guarantee on Crafter}] \label{theorem:emd_converge}
Given dataset pair $(\mathcal{X}_{\text{pvt}}, \mathcal{X}_{\text{pub}})$, Crafter's setup, utility loss $\mathcal{L}_u = l$, we let the optimal attainable PII at $l$ be $\epsilon^*$. With probability at least $1-2\delta$ over the randomness of training samples, we have
\begin{equation}
	\label{theorem:guarantee}
	\begin{split}
		\epsilon - \epsilon^* \leq 4 \mathcal{R}(\mathcal{F}_D)&+4 \mathcal{R}\left(\mathcal{H}_{F_{X} \times D}\right)\\
		&+2\left(Q_{\mathrm{avg}}+Q\right) \sqrt{\frac{\log (1 / \delta)}{2 b s}}.
	\end{split}
\end{equation}
\end{theorem}
The left-hand-side of Eq. \eqref{theorem:guarantee} captures the performance gap between the Crafter generated feature and the ideal one at the same utility loss. Theorem \ref{theorem:emd_converge} states that towards approximating a public prior distribution, Crafter generates feature that approaches the theoretically optimal privacy-utility tradeoff with a bounded distance.

\begin{proof}
First we make the following standard assumptions: the discriminator function class $\mathcal{F}_{\mathcal{D}}$ is defined on compact parameter sets and define  $L_w(i), M_{\omega}(i)$ and $ B_{\text{inv}}$:
\begin{equation}
	\begin{split}
		D\in ~\mathcal{F}_{\mathcal{D}}&:= \{X:||X||\leq B_{\text{avg}}\} \mapsto\\
		&\mathbf{w}_{d}^{\top} \sigma_{d-1}\left(\mathbf{W}_{d-1} \sigma_{d-2}\left(\cdots \sigma_{1}\left(\mathbf{W}_{1} \mathbf{x}\right)\right)\right) \in \mathbb{R},
	\end{split}
\end{equation}
where $\sigma_i(\cdot)$ is $L_w(i)$-Lipschitz continuous activation function of each discrinimator layer $i=1,\dots,d-1$, and $W_i$ are parameter matrices satisfying $||W_i||_F\leq M_{\omega}(i)$. Similarly, we also assume that given feature $F_X$, the corresponding reconstructed images satisfy $||\hat{X}^*||\leq B_{\text{inv}}.$

We define the following abbreviations for the ease of notation. Given feature $F_X$, the white-box attacker $\mathcal{A}$ starts the inversion in Eq.~\eqref{eq:za-loss} from the random initial $z_0$.
We denote the distribution of reconstructed images $G(z^*(F_X))$ as $p_{(F_X;z_0)}$, and denote the prior distrbution $G(z_r)$ as  $p_{\text{avg}}$.
In addition, we use $\hat{p}_{(F_X;z_0)}$ and $\hat{p}_{\text{avg}}$ to denote the empirical distributions of $p_{(F_X;z_0)}$ and $p_{\text{avg}}$ over the samples.

By definition, from $F^*_X$ satisfying $\epsilon$-PII, we have
\begin{equation}
	{\epsilon = \text{EMD}(G(z^*(F^*_X))|| G(z_r)) = d_{nn}(p_{(\vstar; \z)}, \px).}
\end{equation}
Recall that given $X$, $F_X^*$ denotes the Crafter-generated feature via minimizing objective \eqref{eq:ultimate} over the distribution samples, i.e. 
\begin{equation}
F^*_X = \mathop{\arg\min}\limits_{F_X} \text{EMD}(\hat{p}_{(\vstar; \z)}||  \hat{p}_{\text{avg}}) +\beta\cdot \mathcal{L}_u(F_X), 
\end{equation}
from which we have
\begin{equation}
	F^*_X = \mathop{\arg\min}\limits_{F_X\in \{F_X|\mathcal{L}_u(F_X)=l\}}
	d_{nn}(\hat{p}_{(\vstar; \z)}, \hat{p}_{\text{avg}}).
\end{equation}
Also by definition, $\epsilon^*$ is the theoretical optimal attainable PII at utility loss $l$, so
\begin{equation}
	\label{eq:fxdef}
	\epsilon^* = \mathop{\inf}\limits_{F_X\in \{F_X|\mathcal{L}_u(F_X)=l\}}
	d_{nn}(p_{(\vvv; \z)}, \px)
	.
\end{equation}
In the following proof, all features $F_X$ are considered under the $\mathcal{L}_u(F_X)=l$ constraint, and we omit it for the ease of notation.
To this point, we have 

	\begin{equation}
		\begin{aligned}
		\epsilon - \epsilon^*
		=& \dnn( \px, p_{(\vstar; \z)})- \inf_{\vvv} \dnn(\px, p_{(\vvv;\z)}) \\
		=& \underbrace{\dnn( \px, p_{(\vstar; \z)}) - \dnn(\pxhat,p_{(\vstar; \z)})}_\text{(I)}\\
		& + \underbrace{\inf_{\vvv}\dnn(\pxhat, p_{(\vvv; \z)}) - \inf_{\vvv}\dnn(\px, p_{(\vvv; \z)})}_\text{(II)}\\
		&+\underbrace{\dnn(\pxhat,p_{(\vstar;\z)}) -\inf_{\vvv}\dnn(\pxhat,p_{(\vvv;\z)})}_\text{(III)}.
		\end{aligned}
	\end{equation}
	
{We proceed to bound (I), (II) and (III) respectively. Throughout the proof we use this inequality: $\sup x - \sup y\leq\sup(x-y)\leq\sup|x-y|$, which we denote as (*).}

\noindent{\textbf{Bound (I):}.
\begin{equation}
    \label{eq:bound_one}
	\begin{split}
	\text{(I)} =& \sup_D[\mathbb{E}_{X_\text{avg}\sim\px}D(X_\text{avg}) -\mathbb{E}_{{\z}\sim \px}D\circ G(z^*(F_X;\z)))]\\
	&-\sup_D[\mathbb{E}_{X_\text{avg}\sim\pxhat}D(X_\text{avg}) -\mathbb{E}_{{\z}\sim \px}D\circ G(z^*(F_X;\z)))]\\
	\overset{(*)}{\leq}&\sup_D|\mathbb{E}_{X_\text{avg}\sim\px}D(X_\text{avg}) - \mathbb{E}_{X_\text{avg}\sim\pxhat}D(X_\text{avg})|\\
    &=\underbrace{\sup_D|\mathbb{E}_{X_\text{avg}\sim\px}D(X_\text{avg}) - \frac{1}{bs}\textstyle\sum_{i=1}^{b}
    D(X_\text{avg}^{(i)})|}_{R(X_\text{avg}^{(1)}, \dots, X_\text{avg}^{(b)})}.
	\end{split}
\end{equation}
}
\newcommand{\vstaropt}[0]{\tilde{F_X}}

\noindent{\textbf{Bound (II):}}
{
Let $\vstaropt=\arg\min_{F_X}\dnn(\px,p_{(\vvv,\z)})$, and we obtain
\begin{equation}
	\label{eq:proof2}
	\begin{split}
		(\text{II})&\leq  \dnn(\pxhat, p_{(\vstaropt; \z)})-\dnn(\px,p_{(\vstaropt;\z)})\\
		&\overset{(*)}{\leq}
        \underbrace{\sup_D|\mathbb{E}_{X_\text{avg}\sim\px}D(X_\text{avg}) - \mathbb{E}_{X_\text{avg}\sim\pxhat}D(X_\text{avg})|}_{R(X_\text{avg}^{(1)}, \dots, X_\text{avg}^{(b)})  \quad(\text{same as Eq. (\ref{eq:bound_one})})}.
	\end{split}
\end{equation}
}

\noindent{\textbf{Bound (III):}} 
{Let $\hat{F_X}=\arg\min_{F_X}\dnn(\pxhat,p_{(\vvv,\z)})$
\begin{equation}
	\label{eq:proof1}
	\begin{split}
		&\text{(III)}=\dnn(\pxhat,p_{(\vstar;\z)}) - \dnn(\pxhat,\hat{p}_{(\vstar;\z)})\\
		&+\dnn(\pxhat,\hat{p}_{(\vstar;\z)}) - \dnn(\pxhat,p_{(\hat{F_X};\z)})\\
		\overset{\eqref{eq:fxdef}}{\leq}&\dnn(\pxhat,p_{(\vstar;\z)}) - \dnn(\pxhat,\hat{p}_{(\vstar;\z)})\\
		&+\dnn(\pxhat,\hat{p}_{(\hat{F_X};\z)}) - \dnn(\pxhat,p_{(\hat{F_X};\z)})\\
		\leq&\sup_D|\mathbb{E}_{\z\sim\px}D\circ G(z(\vstar;\z)) - \mathbb{E}_{\z\sim\pxhat}D\circ G(z(\vstar;\z))| \\
		&+\sup_D|\mathbb{E}_{\z\sim\px}D\circ G(z(\hat{F_X};\z)) - \mathbb{E}_{\z\sim\pxhat}D\circ G(z(\hat{F_X};\z))|\\
		\leq &2\sup_{D, F_X}|\mathbb{E}_{\z\sim\px}D\circ G(z(F_X;\z)) - \mathbb{E}_{\z\sim\pxhat}D\circ G(z(F_X;\z))|\\
		=&2 \underbrace{\sup_{D, F_X}|\mathbb{E}_{\z\sim\px}D\circ G(z(F_X;\z)) - 
		\frac{1}{bs}\textstyle\sum\limits_{i=1}^{b}D\circ G(z(F_X;\z^{(i)}))|}_{C(z_0^{(1)}, \dots, z_0^{(b)})}.
	\end{split}
\end{equation}
}
{
To upper-bound $C(z_0^{(1)},\dots,z_0^{(b)})$, we observe that $\forall z_0^{(1)}, \dots, z_0^{(1)}, z_0^{(i)'}$, 
\begin{equation}
	\label{eq:proof}
	\begin{split}
		&C(z_0^{(1)}\dots, z_0^{(i)}, \dots, z_0^{(b)}) - C(z_0^{(1)}\dots, z_0^{(i)'}, \dots, z_0^{(b)})\\
		\overset{(*)}{\leq}& \sup_{D, \vvv}|D\circ G(z(\vvv;z_0^{(i)}))-D\circ G(z(\vvv;z_0^{(i)'}))|\\
		\leq& 2Q/b \quad(\text{Cauchy-Schwarz inequality})
	\end{split}
\end{equation}
where \begin{equation}
	Q= B_{\text{inv}}\prod_{i=1}^{d-1} L_{w}(i) \prod_{i=1}^{d} M_{w}(i).
\end{equation}}
{We apply the McDiarmid's inequality on Eq. (\ref{eq:proof}) and obtain with probability at least $1-\delta$:
\begin{equation}
	\begin{split}
		&C(z_0^{(1)}\dots, z_0^{(i)}, \dots, z_0^{(b)}) \\&\leq \underbrace{\mathbb{E}_{z_0}C(z_0^{(1)}\dots, z_0^{(i)}, \dots, z_0^{(b)})}_{\text{(a)}}+2Q\sqrt{\tfrac{\text{log}(1/\delta)}{2b}}.
	\end{split}
\end{equation}}
{
    \begin{equation}
        \begin{split}
            \text{(a)}=&\mathbb{E}_{\z}\sup_D|\mathbb{E}_{\tilde{\z}}\frac{1}{bs}\textstyle\sum_{i=1}^{b} D\circ G(z(\vvv;\tilde{\z}^{(i)}))\\
            &- \frac{1}{bs}\textstyle\sum_{i=1}^{b} D\circ G(z(\vvv;{\z}^{(i)}))|\\
            \leq&\mathbb{E}_{\z, \tilde{\z}^{(i)}}\sup_D|\frac{1}{bs}\textstyle\sum_{i=1}^{b} D\circ G(z(\vvv;\tilde{\z}^{(i)}))\\
            &- \frac{1}{bs}\textstyle\sum_{i=1}^{b} D\circ G(z(\vvv;{\z}^{(i)}))| \quad(\text{Jensen's inequality})\\ 
            =&\mathbb{E}_{\epsilon,\z, \tilde{\z}^{(i)}}\sup_D|\frac{1}{bs}\textstyle\sum_{i=1}^{b} \epsilon_i(D\circ G(z(\vvv;\tilde{\z}^{(i)}))\\
            &-  D\circ G(z(\vvv;{\z}^{(i)})))|\\
            \leq &
            \mathbb{E}_{\epsilon,\z}\sup_D|\frac{1}{bs}\textstyle\sum_{i=1}^{b} \epsilon_i D\circ G(z(\vvv;\tilde{\z}^{(i)}))|\\
            =& 2\mathcal{R}(\mathcal{H}_{\vvv \times D}).
        \end{split}
    \end{equation}
    Therefore,
    \begin{equation}
		\label{eq:proof3}
        \begin{split}
            C(z_0^{(1)}\dots, z_0^{(i)}, \dots, z_0^{(b)})\leq 2\mathcal{R}(\mathcal{H}_{\vvv \times D})+2Q\sqrt{\tfrac{\text{log}(1/\delta)}{2bs}}
        \end{split}
    \end{equation}
    where $Q= B_{\text{inv}}\prod_{i=1}^{d-1} L_{w}(i) \prod_{i=1}^{d} M_{w}(i).$}
{
    Similarly,
    \begin{equation}
		\label{eq:proof4}
        \begin{split}
            R(X_\text{avg}^{(1)}, \dots, X_\text{avg}^{(i)},\dots ,X_\text{avg}^{(b)})\leq 2\mathcal{R}(D)+ 2Q_{\text{avg}}\sqrt{\tfrac{\text{log}(1/\delta)}{2bs}}
        \end{split}
    \end{equation}
    where $Q_{\text{avg}}= B_{\text{avg}}\prod_{i=1}^{d-1} L_{w}(i) \prod_{i=1}^{d} M_{w}(i).$
}

{Summing bounds on (I) (II) and (III), we obtain
 
\begin{equation}
    \begin{split}
        \epsilon - \epsilon^*&=\dnn( \px, p_{(\vstar; \z)})- \inf_{\vvv} \dnn(\px, p_{(\vvv;\z)})\\
        &\leq 4\mathcal{R}(D)+4\mathcal{R}(\mathcal{H}_{\vvv \times D}) +2(Q_{\text{avg}}+Q)\sqrt{\tfrac{\text{log}(1/\delta)}{2bs}}.  
    \end{split}
\end{equation}

}
\end{proof}

\subsection{Validity of $\epsilon$-PII}
\label{sec:emd_appen}
We show that the approximation error between the estimated and theoretical $\epsilon$ is bounded. We denote the estimated value as $\hat{\epsilon}$ and adopt the notations in Appendix \ref{appen:proof}. Then by definition, $\hat{\epsilon} = d_{nn}(\hat{p}_{(\vstar; \z)}, \hat{p}_{\text{avg}})$, and the approximation error is
\begin{equation}
	\begin{aligned}
		&\hat{\epsilon} - \epsilon \\
		=& d_{nn}(\hat{p}_{(\vstar; \z)}, \hat{p}_{\text{avg}}) -
	d_{nn}({p}_{(\vstar; \z)}, {p}_{\text{avg}})\\
	=&d_{nn}(\hat{p}_{(\vstar; \z)}, \hat{p}_{\text{avg}}) - d_{nn}(\hat{p}_{(\vstar; \z)}, {p}_{\text{avg}}) + \\
	& d_{nn}(\hat{p}_{(\vstar; \z)}, {p}_{\text{avg}})
	-d_{nn}({p}_{(\vstar; \z)}, {p}_{\text{avg}})\\
	\overset{(\ref{eq:proof1},\ref{eq:proof2})}{\leq} & C(z_0^{(1)},\dots,z_0^{(b)}) + R(X_\text{avg}^{(1)}, \dots, X_\text{avg}^{(b)})\\
	\overset{(\ref{eq:proof3},\ref{eq:proof4})}{\leq}& 2\mathcal{R}(\mathcal{H}_{\vvv \times D})+2Q\sqrt{\tfrac{\text{log}(1/\delta)}{2b}}
	+2\mathcal{R}(D)+ 2Q_{\text{avg}}\sqrt{\tfrac{\text{log}(1/\delta)}{2b}}
	\end{aligned}
\end{equation}

\section{Design Ideas of Adaptive Attacks}
\label{appen:adaptive_idea}
{  As pointed out by Tramer et al.~\cite{tramer2020adaptive}, no automated tool is able to comprehensively assess a protection's robustness. We thus explore three presently possible adaptive attacks \textbf{A1} to \textbf{A3} that attempt to target the protection's weakest links.
}

\noindent{\textbf{A1: Continue the optimization.}}
A key defence part in Alg.~\ref{alg:opt} is the defender pitting against a simulated worst-case adversary. In the end of the algorithm, $F^*_X$ is released to downstream tasks and will not be updated anymore. As a result, a stronger inversion attack that breaks the previous worst-case assumption may triumph the fixed defence. We design the adaptive adversary to proceed on minimizing a reconstruction loss  $\mathcal{L}_{\text{attacker}}$ (the distance between inverted and original images), which is a consistent loss function. To evaluate our protection comprehensively, besides adjusting the white-box strategy $G$ to optimize the reconstruction loss, we further update black-box $Dec$ as a supplementary adaptive attack.




{A white-box adaptive $\mathcal{A}_1$ tries to enhance its own generator $G$ to match the protected feature $F^*_X$ as below.
It queries Crafter with its images $X\in \mathcal{X}_{\mathrm{adv}}$, intercepts $F^*_{X}$, and reconstructs $G(z^*(F^*_{X}))$, where 
$
z^*(F^*_{X})=\arg\mathop{\min}_{z} \|F^*_{X} - Enc\circ G(z)\|_2
$
}
is the best-reponse of the protected feature.
Then white-box $\mathcal{A}_1$ updates $G$ as
\begin{equation}
\mathop{\min}_{G}\|G(z^*(F^*_{X}))-X\|_2.
\end{equation}
Since the EMD between the reconstructed $G(z^*(F_{X}))$ and average faces $G(z_r)$ is minimized, 
it is equivalent to matching $G(z_r)$ with $X$. As random faces $G(z_r)$ and $X$ are independent and identically distributed image samples, such update only weakens $G$.

{Similarly, a black-box $\mathcal{A}_1$ updates its decoder $Dec$ as 
\begin{equation}
    \mathop{\min}_{Dec}\|Dec(F^*_{X})-X\|_2.
    \end{equation}
}

\noindent{\textbf{A2: Utilize different generators.}}
Another key observation is that the defender's optimization relies on a specific simulated generator model $G$. One may think the defence overfits a particular $G$ {and is not as effective against other adaptive attacks using a different and possibly stronger generator models}. Thereby we evaluate our scheme on generators of different structures and latent dimensions, including the advanced StyleGAN as proposed in~\cite{key}.

\noindent{\textbf{A3: Average features over multiple queries.}} 
{Crafter's protection relies soly on the perturbation on the original features. Therefore, we design this adaptive attack that targets the perturbation defence part.}



\section{Algorithm of Crafter-z}
\label{sec:crafter-z}
Alg. \ref{alg:zp} shows the detailed algorithm of the Crafter-z method that evades implicit differentiation by directly optimizing latent vector $z$. It is the counterpart of Alg.~\ref{alg1}.

In the algorithm, we initialize $z$ as the best-response of $Enc(X)$ so that the corresponding feature representation starts off in the neighborhood of $Enc(X)$ to prevent ineffective utility loss update. $z$ is manipulated to jeopardize discriminator's judgment between the generated distribution and attacker's prior. Meanwhile, the deviation in the feature space is restricted to prevent utility decline. 

\begin{algorithm}[t]
	\renewcommand{\algorithmicrequire}{\textbf{Input:}}
	\renewcommand{\algorithmicensure}{\textbf{Output:}}
	\caption{Crafter-z}
	\begin{algorithmic}[1]
		\label{alg:zp}
		\REQUIRE The same with Alg.~\ref{alg1}.
		\STATE Initialization: $z \leftarrow z^{*}(Enc(X))$, $z_r \leftarrow \texttt{randn}(bs, d)$
		\WHILE{$z$ has not converged}
		\FOR{$t=0,\dots, n_{\mathrm{critic}}$}
		\STATE Sample $\{z^{(j)}\}_{j=1}^m$ a batch from $z.$
		\STATE Sample $\{z_r^{(j)}\}_{j=1}^m$ a batch from random $z_r$.
		\STATE $\mathcal{L}_p \leftarrow 
		\frac{1}{m} \sum_{j=1}^{m} \left[
		D\circ G(z^{(j)})
		-D\circ G(z^{(j)}_{\mathrm{avg}})\right]+gp$
		\STATE $\omega \leftarrow \mathrm{\texttt{AdamOptimizer}}(\nabla_{D}\mathcal{L}_p, D)$
		\ENDFOR
		\STATE $\mathcal{L}_p \leftarrow \frac{1}{bs}\sum_{j=1}^{bs}-D\circ G(z^{(j)})$
		\STATE $v\leftarrow \beta\frac{\partial{\mathcal{L}_u}}{\partial z} + \frac{\partial{\mathcal{L}_p}}{\partial z}$
		\STATE $z \leftarrow \mathrm{\texttt{AdamOptimizer}}(v, z, lr=zlr)$
		\ENDWHILE
		\ENSURE  $Enc\circ G(z^*)$
	\end{algorithmic}  
\end{algorithm}

\section{Attack Hyperparameters}
\label{sec:attacker_param}
We provide the hyperparameters for the white-box, black-box, hybrid white-box attacks and their adaptive versions.

For the model deployment scenario in \S\ref{sec:inference}, white-box attacks perform 600 iterations of optimization for CelebA and LFW respectively. The latent vectors are of dimension 500 for CelebA and 3000 for LFW. In Crafter-z, the learning rate of latent vector is $lr_z = 0.005$ for both datasets. The hybrid white-box attack optimizes over 150 iterations on the image of LFW with a learning rate of $lr_X = 0.005$. 
Adaptive white-box attacks use $G$ of white-box attacks as initial parameters, and update $G$ using RMSProp optimizer on $\mathcal{X}_{\mathrm{pub}}$ and the corresponding $z_{inv}$ (calculated by inversion) pairs with $lr =0.001$.  Black-box attacks trains in total 500 epochs using Adam optimizer with default hyperparameters defined in PyTorch: $lr=0.001$, $betas=(0.9, 0.999)$, $eps=10^{-8}$, $weight\_decay=0$.  Adaptive black-box attacks use the black-box attack $Dec$ as the initial point, and update 70 more epochs on the protected feature and image pairs. An Adam optimizer with default hyperparameters is adopted.

For the training scenario, white-box attacks perform 1000 iterations of optimization on latent vector of dimension 700 and $lr_z=0.03$. The update of $G$ of the adaptive white-box attacks uses a RMSProp optimizer with $lr = 0.001$.  Hyperparameters of black-box and adaptive black-box attacks are the same with those in the inference scenario.

\section{Model Architectures}
\label{sec:architecture}
Table \ref{tab:archResnetVGG}, \ref{tab:archD}, \ref{tab:archGen}, \ref{tab:archAmor} respectively show the target networks, the discriminator, the generator, and the amortized network in the experiments.
\begin{table}[h]
	\small
	\centering
	\caption{Architecture of the target models.}
	\begin{tabular}{c|c|c}
		\hline
		& \multicolumn{1}{c|}{ResNet18} & \multicolumn{1}{c}{VGG16} \\ \hline
		\multicolumn{1}{l|}{Image}  & 3$\times$64$\times$64 (3$\times$128$\times$128)           & 3$\times$64$\times$64 (3$\times$128$\times$128)       \\ \hline
		\multirow{2}{*}{$Enc$}    & conv 7$\times$7, 64, stride 2        & (3$\times$3, 64) $\times$2, maxpool      \\
		& $\begin{bmatrix}3\times 3, & 64\\3\times 3,&64\end{bmatrix} \times 2$                    & (3$\times$3, 128) $\times$2, maxpool     \\ \hline
		Feature                     & 64$\times$16$\times$16 (64$\times$32$\times$32)           & 128$\times$16$\times$16 (128$\times$32$\times$32)     \\ \hline
		\multirow{4}{*}{$f$} & $\begin{bmatrix}3\times 3, & 128\\3\times 3,&128\end{bmatrix} \times 2$                    & (3$\times$3, 256) $\times$3, maxpool     \\
		& $\begin{bmatrix}3\times 3, & 256\\3\times 3,&256\end{bmatrix} \times 2$                    & (3$\times$3, 512) $\times$3, maxpool     \\
		& $\begin{bmatrix}3\times 3, & 512\\3\times 3,&512\end{bmatrix} \times 2$                    & (3$\times$3, 512) $\times$3, maxpool     \\
		& avgPool, linear                   & avgPool, linear$\times$3 \\ \hline         
	\end{tabular}
	\label{tab:archResnetVGG}
\end{table}
\begin{table}[h]
	\small
	\centering
	\caption{Architecture of discriminator models.}
	\begin{tabular}{c}
		\hline
		$D$         \\ \hline
		Conv 5$\times$5, 64, stride 2  \\
		Conv 5$\times$5, 128, stride 2 \\
		Conv 5$\times$5, 256, stride 2 \\
		Conv 5$\times$5, 512, stride 2 \\
		Conv 4$\times$4, 1, stride 1   \\ \hline
	\end{tabular}
	\label{tab:archD}
\end{table}
\begin{table}[h]
	\small
	\centering
	\caption{Architecture of generator models.}
	\begin{tabular}{c|c}
		\hline

		$G_1$                          & $G_2$          \\ \hline
		Linear(input dim, 64$\times$256) & Linear(input dim, 64$\times$128) \\
		BatchNorm+ReLU             & BatchNorm+ReLU             \\
		Reshape(64$\times$4,8,8)          & Reshape(64$\times$8,4,4)          \\
		ConvTranspose(64$\times$4, 64$\times$2)  & ConvTranspose(64$\times$8, 64$\times$4)  \\
		ConvTranspose(64$\times$2, 64)    & ConvTranspose(64$\times$4, 64$\times$2)  \\
		ConvTranspose(64, 3)       & ConvTranspose(64$\times$2, 64)    \\
		& ConvTranspose(64, 3)     \\
		\hline 
	\end{tabular}
	\label{tab:archGen}
\end{table}
\begin{table}[h]
	\centering
	\small
	\caption{Architecture of amortize net, omitting batch norm layers and leaky ReLU after each convolution.}
	\begin{tabular}{c}
		\hline
		Amortize Net             \\ \hline
		Conv 4$\times$4 128 stride 2    \\
		Conv 4$\times$4 512 stride 2    \\
		Conv 4$\times$4 1024 stride 2   \\
		Conv 2$\times$2 512 stride 1    \\
		Conv 1$\times$1 dim\_z stride 1 \\ \hline
	\end{tabular}
	\label{tab:archAmor}
\end{table}

\section{Inference Results: FSIM vs. Utility}
\subsection{Against Black and White-Box Attacks}
\label{sec:FSIM}
Figure \ref{fig:fsimBlack} and \ref{fig:fsimWhite} shows the FSIM-utility tradeoff against black-box and white-box attack respectively. Consistent with the conclusion in \S\ref{sec:inference}, Crafter achieves the most desirable utility-privacy tradeoff.
\begin{figure}[t]
	\centering
	\subfigure[]{
		\centering
		\includegraphics[width=0.5\linewidth]{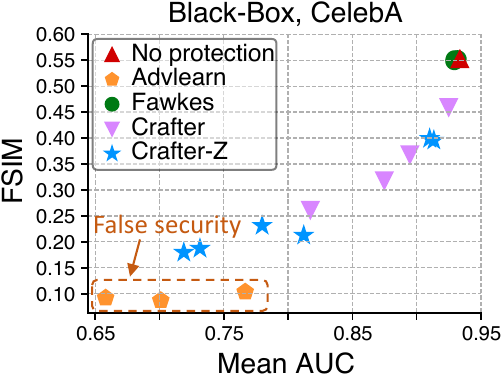}
	}%
	\subfigure[]{
		\centering
		\includegraphics[width=0.5\linewidth]{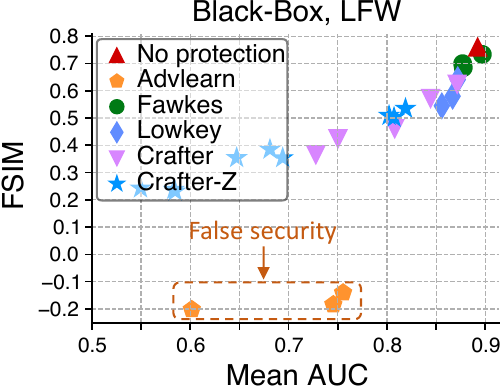}
	}

	\caption{Privacy-utility tradeoffs against black-box attacks. On CelebA, $\beta \in \{0.5,1,2,10\}$ for Crafter, and $\beta \in \{20, 50\}$ for Crafter-z. On LFW, $\beta\in \{3.5, 4, 4.5, 6, 7\}$ for Crafter, and $\beta\in \{5, 10, 20\}$ for Crafter-z.}
	\label{fig:fsimBlack}
\end{figure}
\begin{figure}[t]
	\centering
	\subfigure[]{
		\centering
		\includegraphics[width=0.5\linewidth]{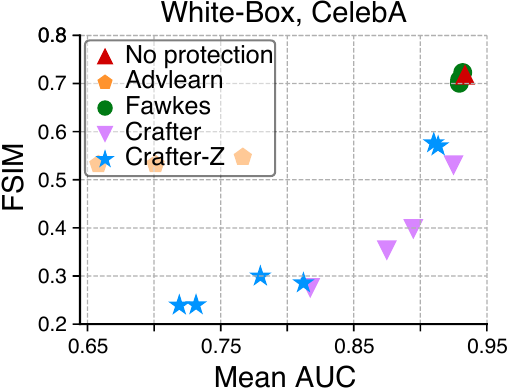}
	}%
	\subfigure[]{
		\centering
		\includegraphics[width=0.5\linewidth]{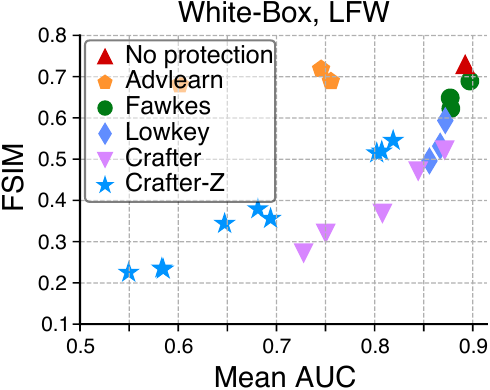}
	}

	\caption{Privacy-utility tradeoffs against white-box attacks. On CelebA, $\beta \in \{0.5,1,2,10\}$ for Crafter, and $\beta \in \{20, 50\}$ for Crafter-z. On LFW, $\beta\in \{3.5, 4, 4.5, 6, 7\}$ for Crafter, and $\beta\in \{5, 10, 20\}$ for Crafter-z.}
	\label{fig:fsimWhite}
\end{figure}

\subsection{Against Hybrid White-Box Attacks}
\label{sec:hybridFSIM}
Figure \ref{fig:fsimHybrid} shows the FSIM-utility tradeoff against hybrid white-box attack. Consistent with the conclusion in \S\ref{sec:inference}, adversarial learning algorithm reaches an extremely low FSIM, through an adversarial-example-like way.

\begin{figure}[t]
	\centering
	\includegraphics[width=0.5\linewidth]{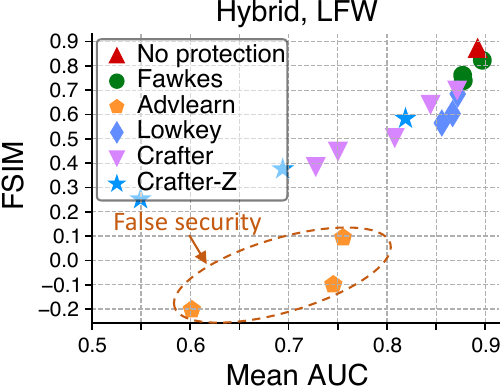}	
	
	\caption{FSIM-utility tradeoffs against hybrid white-box  attacks.}
	\label{fig:fsimHybrid}
\end{figure}

\section{Inference Results: Adaptive Attacks}
\label{sec:appendix_adapt}
Figure \ref{fig:fsimAdaptiveWhite} and Figure \ref{fig:fsimAdaptiveBlack} show respectively the changes of FSIM as the adaptive white-box attack and adaptive black-box attack training goes. Figure \ref{fig:adaptiveRest} is a supplement of Figure \ref{fig:adaptive_black}.
\begin{figure}[t]
	\centering
	\includegraphics[width=\linewidth]{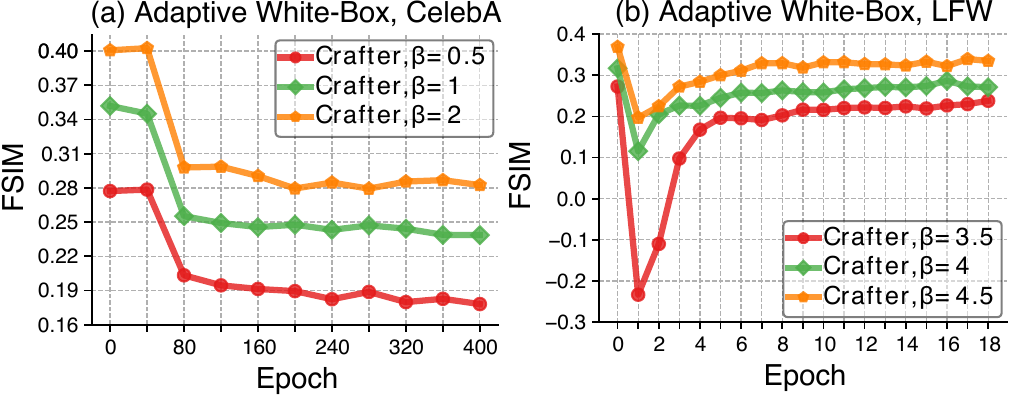}	
	\caption{FSIM changes along training epochs of adaptive white-box attack.}
	\label{fig:fsimAdaptiveWhite}
\end{figure}
\begin{figure}[t]
	\centering
	\includegraphics[width=\linewidth]{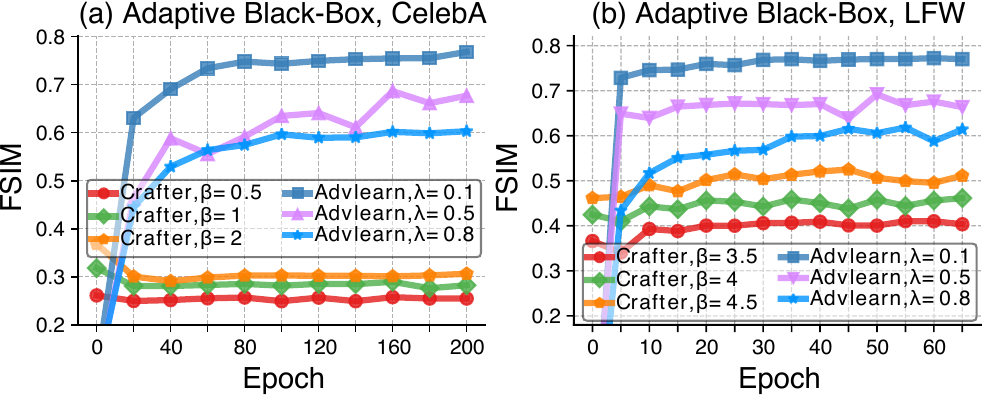}	
	\caption{FSIM changes along training epochs of adaptive black-box attack.}
	\label{fig:fsimAdaptiveBlack}
\end{figure}
\begin{figure}[t]
	\centering
	\includegraphics[width=\linewidth]{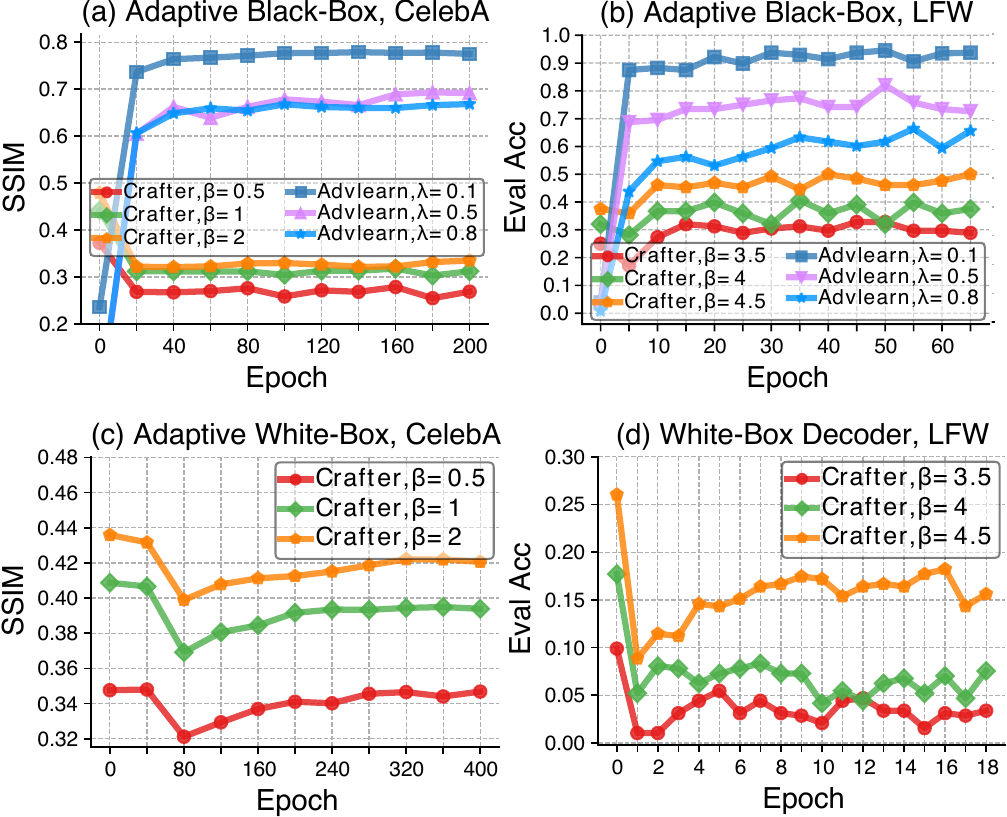}	
	\caption{Crafter and Adv Learning on CelebA and LFW against adaptive black/white box attacks, as a supplement of Figure \ref{fig:adaptive_black}.}
	\label{fig:adaptiveRest}
\end{figure}




\section{Inference Results:  Azure Accuracy}  
\label{sec:append_inference}
Figure \ref{fig:azure} shows the Azure Eval Acc and utility tradeoff. The tradeoff is consistent with those in Figure~\ref{fig:scatter_white}. Therefore, Crafter is indeed effective against white-box reconstruction attacks when evaluated by a commercial face verification API.
\begin{figure}[htbp]
	\centering
	\includegraphics[width=0.7\linewidth]{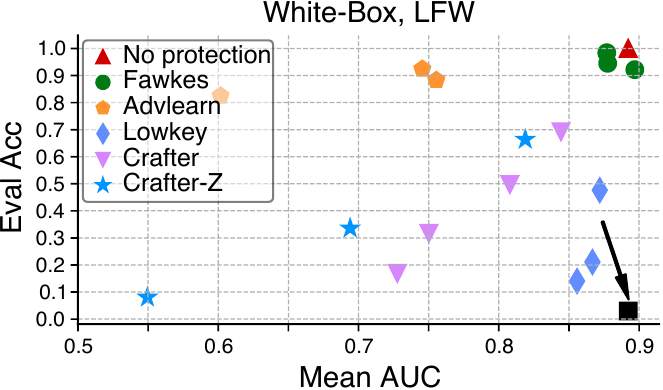}
	\caption{Azure Eval Acc under white-box attacks, on LFW.
	}
	\label{fig:azure}
\end{figure} 

\section{Defects of Crafter-z}
\label{sec:crfz}
Crafter-z is not ideal in the sense that it usually offers poor tradeoffs which are difficult to manipulate. We first found out feasible ranges of $\beta$ and $lr_{z}$ to achieve acceptable privacy and utility in Crafter-z. Table~\ref{tab:zp} demonstrates the impact of $\beta$ and $lr_{z}$ of Crafter-z on CelebA. For fixed $lr_{z}$, changing $\beta$ from 20 to 50 does not affect the Eval Acc and AUC much. But for fixed $\beta$, varying $lr_{z}$ deliver a clear tradeoff. This drawback is more prominent on LFW with larger hyperparameter sets: marked yellow are tradeoff points of Crafter-z appearing in clusters of three. Points within each cluster share the same $lr_{z}$ with $\beta=5, 10, 20$. This phenomenon again illustrates that $lr_{z}$ dominates the tradeoff rather than the expected $\beta$. In addition to poor tradeoff, this undesirable behavior is a second reason why Crafter-z is not preferred.
\begin{table}[htbp]
    \small
    \centering
    \caption{Crafter-z tradeoff on CelebA, inference scenario.}
    \begin{tabular}{cc|cc|c}
    \hline
    \multicolumn{2}{c|}{Hyperparams} & \multicolumn{2}{c|}{Eval Acc \%} & AUC     \\ \hline
    $lr_{z}$                  & $\beta$                  & white-box           & black-box        &    mean AUC         \\\hline
    0.0001               & 20                    & 34.11    & 10.16    & 91.33 \\
    0.0001               & 50                    & 35.94       & 10.94     & 91.01 \\\hline
    0.0005               & 20                    & 9.89   & 7.03    & 77.98  \\
    0.0005               & 50                    & 7.81       & 3.13      & 81.22 \\\hline
    0.001                & 20                    & 5.73    & 2.34    & 71.89 \\
    0.001                & 50                    & 2.86   & 2.34   & 73.15 \\ \hline
    \end{tabular}
    \label{tab:zp}
    \end{table}

\section{Speed-up with Amortizer}
\label{sec:amor}
The goal of amortizer is to speed up the inversion in line 3 of Alg.~\ref{alg1} by establishing a mapping from feature space to latent space. It receives a batch of feature as input, and is expected to output a latent vector that approximates the real best-response computed via optimization. The white-box attacker uses the output latent vector of the amortizer as its initial point in the optimization, thereby cutting off the time expense.

Specifically, to train the amortizer, we generate some latent vector $z$ following a random distribution, calculate the corresponding features $Enc(G(z))
$ and optimize the amortizer $Amor$ as follows:
\begin{equation}
	\mathop{\min}_{Amor} (z, Amor\circ Enc \circ G(z))).
\end{equation}
The structure of amortizer we adopt is listed in Table \ref{tab:archAmor}.
We refer to a white-box inversion with the assistance of amortizer as the \emph{amortized inversion}.
Empirically, we verify that on the LFW dataset, it takes only 100 iterations for an amortized inversion to achieve the same level of reconstruction performance with that reached by 600 normal inversion iterations, thereby greatly reducing the computation overhead.

\section{{Visualization}}
\label{sec:vis}
Figure~\ref{fig:heat_celeb} { shows the heatmap of CelebA.}
Figure \ref{fig:vis2}, \ref{fig:vis3} are supplement visualizations for Figure \ref{fig:heatvis} to illustrate the effectiveness of different protection schemes against various classes of attacks on LFW, $\lambda = 0.5, \beta=4.5$.
These visualization results further confirm the conclusion as stated in~\ref{sec:adaptive}.

\begin{figure}[h]
    \centering
    \includegraphics[width=2.8in]{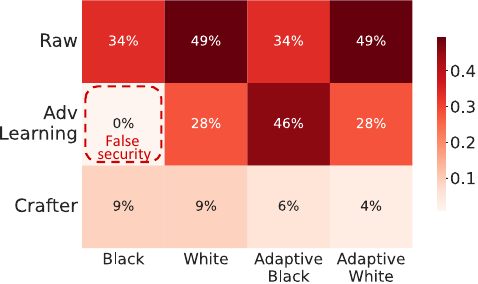}
    \caption{Average Eval Acc of protection schemes {on CelebA} against different attacks, deployment scenario, averaged across $\beta\in\{0.5, 1, 2\}$.}     
    \label{fig:heat_celeb}
\end{figure}

{Figure \ref{fig:lowkey_images} shows the reconstruction results under LowKey protection. Regardless of the protection mode, the reconstructed images resembles the original image, and the protection is ineffective.}


\begin{figure}[h]
    \centering
    \includegraphics[width=0.9\linewidth]{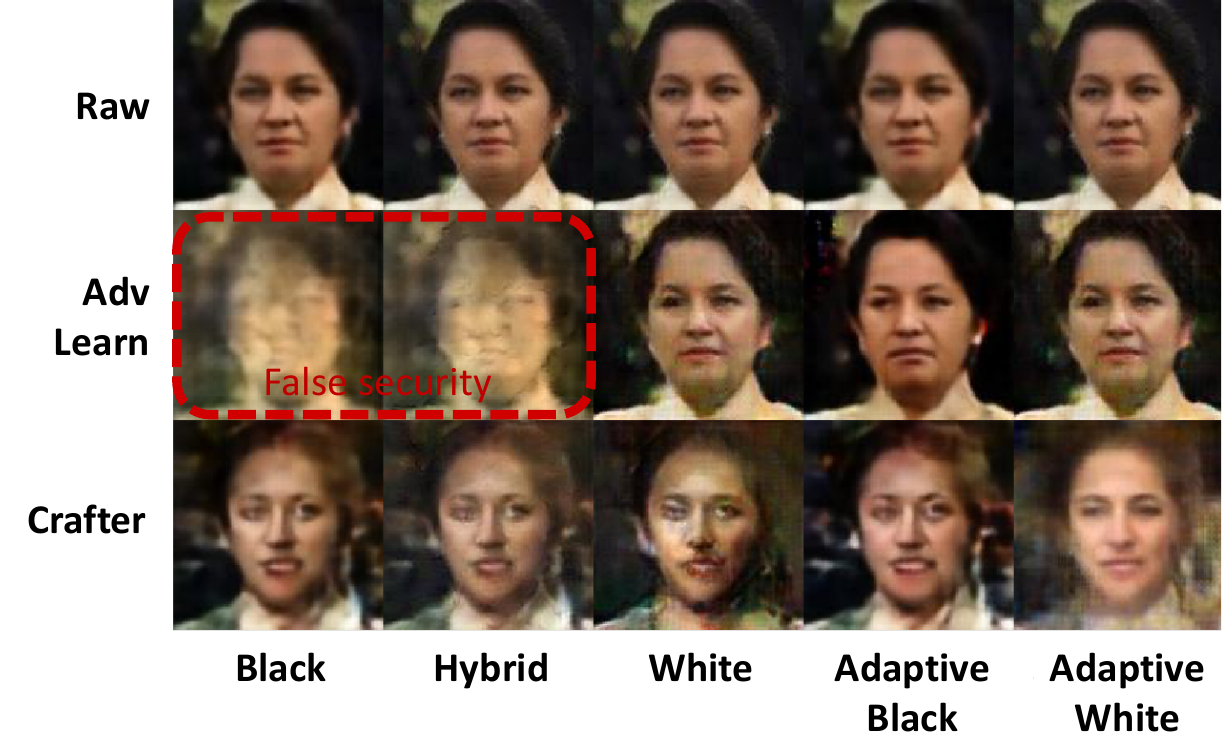}
    \caption{A visualization effect of protection schemes against different attacks. $\lambda = 0.5, \beta=4.5$.}
    \label{fig:heatvis}     
\end{figure}
\begin{figure}[h]
	\centering
	\includegraphics[width=0.9\linewidth]{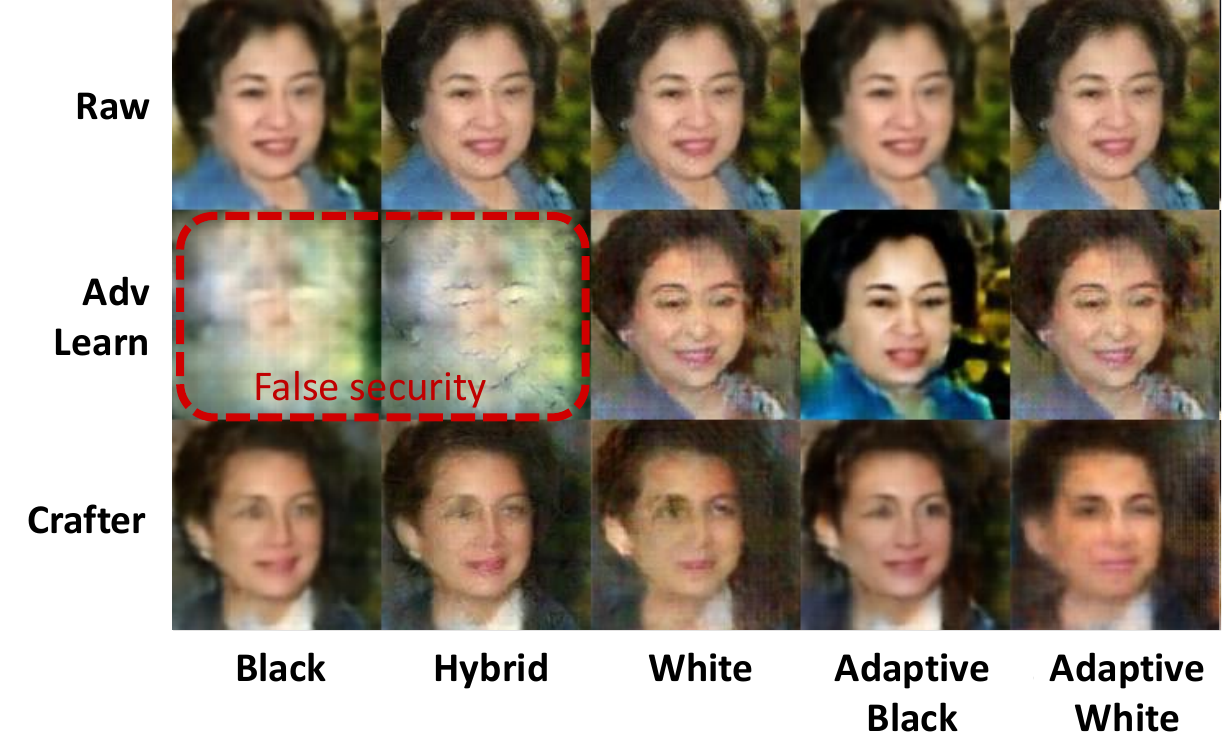}
	\caption{Visualization results for different protections against different attacks.}
	\label{fig:vis2}
\end{figure}

\begin{figure}[h]
	\centering
	\includegraphics[width=0.9\linewidth]{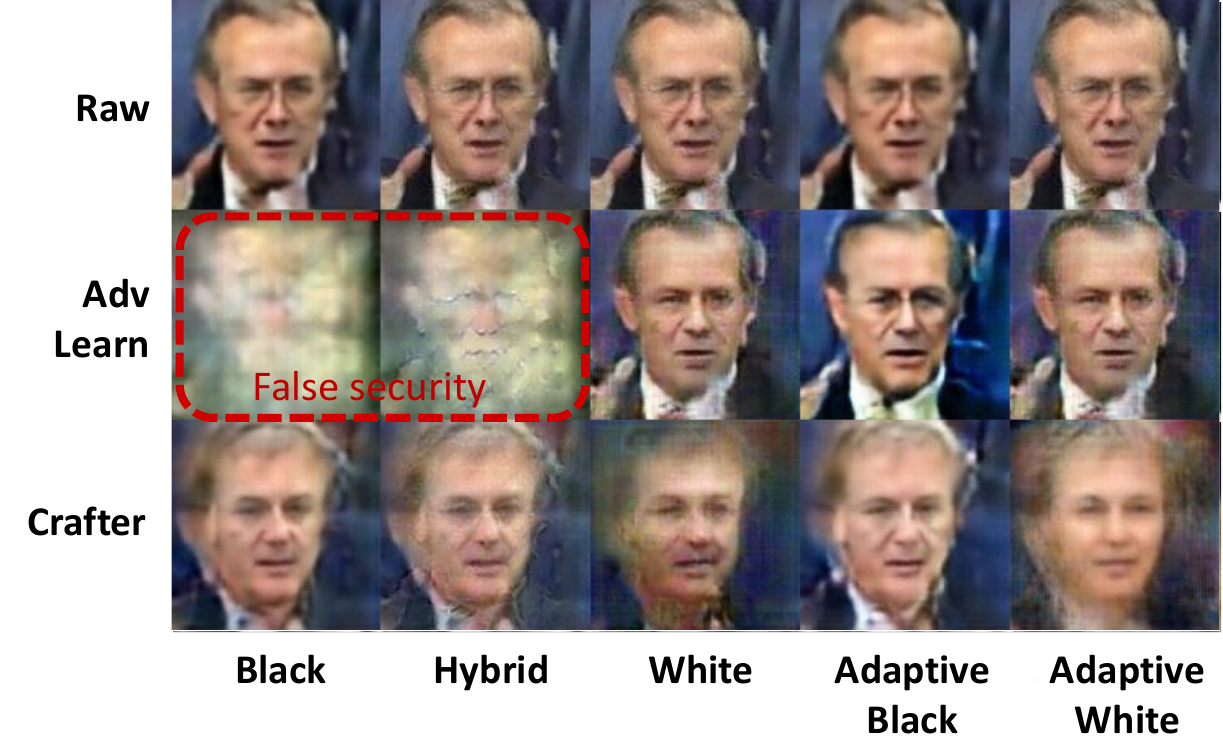}
	\caption{More visualization results.}
	\label{fig:vis3}
\end{figure}

\section{Human Study Detail}
\label{appen:human}
Details of the human study protocol are as follows. 
The volunteers are recruited from social media open to university students. Based on the statistics of the poll, 29\% of participants are female, all are under 30 ages and all are Asian. We pay 0.5\$ for each test.
The poll consists of 10 questions, each with a different identity, and two of the questions do not have the correct answers among the provided options. 


We eventually harvest 35 valid feedbacks each from a different individual  and 
compute the poll statistics from the harvested feedbacks as in Table \ref{tab:human2}.

\begin{table}[ht]
		\small
		\centering
		\caption{Human study statistics.}
		\begin{tabular}{c|c|c|c|c|c|c}
		\hline
		 & A & B&C&D&`None' & Macro-F1 score    \\ \hline
		Recall               & 0.100   & 0.190    & 0.357   & 0.343 & 0.386 \\
		Precision              & 0.115                  & 0.114      & 0.347    & 0.363 & 0.248 \\\hline
		F1              & 0.107& 0.143  & 0.352 & 0.353 & 0.302 & 0.251 \\\hline
		\end{tabular}
		\label{tab:human2}
\end{table}


\end{document}